\documentclass[lettersize,journal]{IEEEtran}
\usepackage[dvipdfmx]{graphicx}
\usepackage{amsmath,amsfonts}
\usepackage{amssymb}
\usepackage{algorithmic}
\usepackage{algorithm}
\usepackage{array}
\usepackage{textcomp}
\usepackage{stfloats}
\usepackage{url}
\usepackage{verbatim}
\usepackage{graphicx}
\usepackage{cite}
\usepackage[acronym,shortcuts]{glossaries}
\usepackage[dvipsnames]{xcolor}
\usepackage{color}
\usepackage[font=footnotesize]{subcaption}
\usepackage[font=footnotesize]{caption}

\definecolor{MRIblue}{rgb}{0.008, 0.227, 0.502}
\definecolor{MRIred}{rgb}{0.8627, 0, 0.2353}
\definecolor{MRIyellow}{rgb}{0.8627, 0.6275, 0}
\definecolor{MRIgreen}{rgb}{0.1961, 0.6078, 0.4510}
\usepackage{eso-pic}
\usepackage{tikz}

\AddToShipoutPicture{%
    \AtPageLowerLeft{%
        \ifnum\value{page}=1 % 1ページ目のみ適用
        \begin{tikzpicture}[remember picture, overlay]
            \node[draw, thick, rectangle, fill=white, text width=16cm, align=center]
                at (10.5cm,0.8cm) {\textbf{\textsf{This work has been submitted to the IEEE for possible publication. Copyright may be transferred without notice, after which this version may no longer be accessible.}}};
        \end{tikzpicture}
        \fi
    }
}
\usepackage[colorlinks=true,citecolor=blue,linkcolor=MRIred]{hyperref}

\newacronym{UE}{UE}{user equipment}
\newacronym{BS}{BS}{base station}
\newacronym{MIMO}{MIMO}{multiple-input-multiple-output}
\newacronym{MISO}{MISO}{multiple-input-single-output}
\newacronym{LoS}{LoS}{line-of-sight}
\newacronym{NLoS}{NLoS}{non-line-of-sight}
\newacronym{ULA}{ULA}{uniform linear array}
\newacronym{SNR}{SNR}{signal-to-noise ratio}
\newacronym{AoA}{AoA}{angle of arrival}
\newacronym{AoD}{AoD}{angle of departure}
\newacronym{mmWave}{mmWave}{millimeter-wave}
\newacronym{AWGN}{AWGN}{additive white Gaussian noise}
\newacronym{RF}{RF}{radio-frequency}
\newacronym{ADC}{ADC}{analog-to-digital converter}
\newacronym{TDD}{TDD}{time-division duplexing}
\newacronym{UL}{UL}{uplink}
\newacronym{DL}{DL}{downlink}
\newacronym{CSI}{CSI}{channel state information}
\newacronym{HPBW}{HPBW}{half power beamwidth}
\newacronym{CDF}{CDF}{cumulative distribution function}
\newacronym{PDF}{PDF}{probability density function}
\newacronym{PMF}{PMF}{probability mass function}
\newacronym{SE}{SE}{spectral efficiency}
\newacronym{DFT}{DFT}{discrete Fourier transformation}
\newacronym{IIoT}{IIoT}{industrial internet of things}
\newacronym{B5G}{B5G}{beyond fifth-generation mobile communications systems}
\newacronym{6G}{6G}{sixth-generation mobile communications systems}
\newacronym{RSNR}{RSNR}{received signal-to-noise ratio}
\newacronym{CoMP}{CoMP}{coordinated multi-point}
\newacronym{GA}{GA}{genetic algorithm}
\newacronym{XL-MIMO}{XL-MIMO}{Extremely large-scale MIMO}
\newacronym{NMSE}{NMSE}{normalized mean square error}
\newacronym{GPU}{GPU}{graphics processing unit}
\newacronym{CS}{CS}{compressed sensing}
\newacronym{3GPP}{3GPP}{3rd Generation Partnership Project}

\newcommand{\figref}[1]{\figurename~\ref{#1}}
\newcommand{\algref}[1]{Alg.~\ref{#1}}
\newcommand{\appref}[1]{Appendix-\ref{#1}}

\begin{document}
\sloppy
\title{Revisiting Beamforming Design for Stable Millimeter-Wave Communications Under Blockages}

\author{Kanta Terui,~\IEEEmembership{Graduate Student Member, IEEE}, Kabuto Arai,~\IEEEmembership{Member, IEEE}, \\and Koji Ishibashi,~\IEEEmembership{Senior Member, IEEE\\
\vspace{-5ex}}
% <-this % stops a space
\thanks{This work was supported by KDDI Research, Inc.
An earlier version of this paper was presented in part at the 2024 IEEE 100th Vehicular Technology Conference (VTC2024-Fall) [DOI: 10.1109/VTC2024-Fall63153.2024.10757745].}% <-this % stops a space
% \thanks{Copyright (c) 2015 IEEE. Personal use of this material is permitted. However, permission to use this material for any other purposes must be obtained from the IEEE by sending a request to pubs-permissions@ieee.org.}%
\thanks{K. Terui and K. Ishibashi are with the Advanced Wireless and Communication Research Center (AWCC), The University of Electro-Communications, Tokyo 182-8285, Japan (e-mail: terui@awcc.uec.ac.jp,  koji@ieee.org)}
\thanks{K. Arai is with the Graduate School of Science and Engineering, Saitama University,  Saitama 338-8570, Japan (e-mail: kabuto@ieee.org)}
}

\markboth{Journal of \LaTeX\ Class Files,~Vol.~XX, No.~X, XXXX~2026}%
{Shell \MakeLowercase{\textit{et al.}}: A Sample Article Using IEEEtran.cls for IEEE Journals}

% % column for its text to clear the IEEEpubid mark.

\maketitle
\begin{abstract}
    This paper investigates robust analog beamforming for millimeter-wave (mmWave) communications under stochastic path blockages using multi-panel arrays.
    Conventional designs concentrate beams on the line-of-sight (LoS) path to maximize array gain, but this approach is highly vulnerable to sudden disconnections when the LoS path is blocked.
    To overcome this limitation, we propose a multi-beam design that exploits both LoS and non-line-of-sight (NLoS) paths for stable communications.
    The major contribution of this work is to establish a theoretical foundation for multi-beam design, where closed-form expressions for the cumulative distribution function (CDF) and outage probability of the spectral efficiency (SE) are derived.
    To design the optimal multi-beam based on the derived outage probability, we formulate a panel allocation problem to determine the assignment of panels to specific paths.
    The optimization problem can be solved by two algorithms based on brute-force search.
    Through computer simulations, the validity of the theoretical analysis for multi-beam design is confirmed, and the proposed algorithms substantially reduce the outage probability while maintaining average SE performance, thereby achieving stable communication.
\end{abstract}


\glsresetall

\section{Introduction}
\IEEEPARstart{T}{he} demand for wireless communication has seen unprecedented growth with the advent of \ac{B5G} and \ac{6G} systems \cite{sixkey_heng}.
These systems are envisioned to support a massive increase in data traffic while enabling ubiquitous connectivity among devices and users.
For instance, applications such as autonomous vehicle control, remote medical procedures, and \ac{IIoT} require the stable exchange of large volumes of data, including high-definition video streams \cite{6G_NeiKato,6G_martin,6G_Chowdhury}.
Achieving stable, high data rate communication under such conditions is critical to fulfilling these demands.
Among the various approaches, the use of \ac{mmWave} frequency bands has garnered significant attention due to its potential to provide wider bandwidths, thereby enhancing throughput.
However, mmWave communications face challenges such as severe path loss caused by water vapor absorption and signal attenuation due to reflection and diffraction, leading to significant degradation in received power \cite{pathloss_rappaport}.
To mitigate these challenges, the inherent short wavelength of mmWave signals enables the deployment of large-scale antenna arrays capable of implementing high-gain beamforming to compensate for path loss\cite{overview_heath,mMIMO_larsson}.

\par
In practical systems, large-scale arrays are often implemented as multi-panel arrays where each panel is an integrated circuit that includes power amplifiers, a limited number of antenna elements, and the corresponding phase shifters~\cite{overview_heath,mp_overview,mp_song,mptruncated_wang,mp_3GPP}.
This approach reduces circuit design costs and minimizes power consumption, allowing for industrial deployment.
Furthermore, to address the cost associated with \ac{ADC} design, the antenna arrays are commonly adopted with a single \ac{RF} chain, requiring analog beamforming techniques.

\par
Current beamforming designs for \ac{mmWave} systems often rely on codebook-based approaches, where predesigned codebooks are used to select beams that align with the \ac{LoS} path between a \ac{BS} and a \ac{UE} to acquire high array gain ~\cite{3GPP_SSB,3GPP_LSP,3GPP_FDMIMO}.
While this approach is effective for maximizing received power under static channel conditions, practical mmWave channels are highly susceptible to sudden path blockages caused by obstacles such as pedestrians, vehicles, or falling leaves~\cite{humanblockage_mukherjee,empirical_slezak,rapid_maccartney,blockage_modeling_akdeniz,pathloss_rappaport,pblk_raghavan,blockage_li}.
Such dynamic changes due to blockages cause beam misalignment for the selected beam from the codebook, which is aligned to only the \ac{LoS} path, resulting in an unstable communication system with frequent disconnections.
The frequent reconfiguration of beams to reestablish the communication link requires additional overhead, which hinders the overall system performance.

\par 
Numerous studies have addressed the issue of mitigating the impact of path blockage in \ac{mmWave} communications \cite{blockage_iimori,blockage_uchimura,blockagehybrid_uchimura,blockagewcnc_uchimura,NLoS_Dogan,mmWaveReliable_Kumar,BeamSwitch_Wilson,mp_terui,latency_kumar,blockage_kumar,blockage_nishio}.
To realize stable communication under blockage environments, the authors in~\cite{blockage_iimori,blockage_uchimura,blockagehybrid_uchimura,blockagewcnc_uchimura} proposed an optimal beamforming design that minimizes the outage probability of the \ac{SE} in \ac{CoMP} systems employing multiple \acp{BS}.
However, directly applying these methods to practical mmWave systems is challenging, because these studies assume that analog beams can be flexibly designed with only an equal-amplitude constraint, without relying on the predefined codebook-based beams commonly used in commercial systems.
Moreover, computational complexity poses a significant challenge, because these methods require calculating the outage probability with the Monte Carlo method and solving the optimization problem at every channel coherence time based on instantaneous \ac{CSI}.

To focus on practical mmWave systems under blockage, the authors in \cite{NLoS_Dogan,mmWaveReliable_Kumar,BeamSwitch_Wilson} investigate the effects of path blockage and beamforming design to address the issue in practical systems.
These works claim that leveraging multiple paths, including not only the \ac{LoS} path but also \ac{NLoS} paths, can achieve spatial diversity and enhance communication stability in the presence of path blockage.
The works in \cite{mmWaveReliable_Kumar,BeamSwitch_Wilson} have proposed multi-beam design where the designed multiple beams are aligned to both the LoS and the NLoS path to achieve spatial diversity.
The effectiveness of leveraging multiple paths through the multi-beam design is validated numerically and experimentally, demonstrating the feasibility of achieving stable communication systems.
However, the evaluations in these works are limited in simple scenarios with only two paths including a LoS path and a single NLoS path, for the sake of simplifying the performance evaluation and multi-beam design.
To further extend multi-beam design in more practical scenarios, the authors in \cite{mp_terui} proposed a multi-beam design with multi-panel arrays relying on codebook-based beams.
By independently controlling each panel in a multi-panel array, multiple beams aligned to both the \ac{LoS} path and multiple \ac{NLoS} paths can be designed, reducing the outage probability and leading to stable communication.
However, this method uses a naive multi-beam design with equal panel allocation for each path, without considering outage probability optimization, and lacks theoretical analysis for the multi-beam design.

Since the prior studies~\cite{NLoS_Dogan,mmWaveReliable_Kumar,BeamSwitch_Wilson,mp_terui} are limited to simple multi-beam design without a supporting theoretical analysis, this paper establishes a theoretical framework for multi-beam design with a multi-panel array. 
The objective is to ensure stable communication under blockage environments by minimizing the outage probability of the \ac{SE}.
Furthermore, while AI/ML-based approaches have also been investigated for proactive blockage or received-power prediction \cite{blockage_nishio}, they generally require offline data collection and model training.
In contrast, the proposed method provides a lightweight, training-free alternative based on closed form outage analysis and long-term channel statistics.
Unlike conventional approaches for outage minimization in \ac{CoMP} ~\cite{blockage_iimori,blockage_uchimura,blockagehybrid_uchimura,blockagewcnc_uchimura}, our work investigates a method for mitigating the impact of path blockage with a multi-panel array under the practical constraint of a simple single-\ac{BS} configuration that does not require additional hardware or signaling.
In addition, we propose two algorithms to solve the outage minimization problem based on the theoretical analysis. 
The main contributions of this paper are summarized as follows.
\begin{itemize}
    \item
        \textbf{Theoretical analysis of the SE}:
        The CDF and outage probability of the SE, when using the multi-beam design with the multi-panel array, is derived in closed form.
        Based on an approximation of array responses~\cite{sinc_wang}, the equivalent channel, including analog beams and blockage effects, can be simplified for analytical tractability.
        Owing to the approximation, the distribution of the equivalent channel can be represented as a Bernoulli-Gaussian mixture distribution, which is composed of the weighted summation of a delta function and multiple Gaussian distributions, where the variance of each distribution depends on the beamforming and blockage effects.
        Using this distribution, the outage probability of the SE and \ac{RSNR} can be derived in closed form.
        The closed-form expression of the outage probability enables theoretical discussion and mathematical formulation for the optimization of multi-beam design, addressing how the multiple beams should be aligned to each path in order to minimize the outage probability under blockage environments.
    \item
        \textbf{Optimization for panel allocation}: 
        Based on the derived outage probability, we propose two analog beamforming designs: one that considers only the minimization of the outage probability and another that considers both the outage probability and the average SE. 
        To design the optimal multi-beam, we formulate a panel allocation problem to determine the allocation of panels to specific paths, including both LoS and NLoS paths. Although the panel allocation problem is formulated as a combinatorial optimization, it is feasible to solve the problem with reasonable complexity even when using brute-force search, which can obtain global optima.
        This is because the parameters affecting the complexity such as the number of panels and the number of paths are relatively small values in practical mmWave systems.
        Unlike the conventional approaches~\cite{blockage_iimori,blockage_uchimura,blockagehybrid_uchimura,blockagewcnc_uchimura}, we can derive the outage probability in closed form without Monte Carlo computation, which reduces the computational complexity for analog beamforming design. 
        Moreover, the proposed algorithm requires only long-term statistics, such as the number of paths and the Rician $K$-factor, rather than instantaneous \ac{CSI}. 
        Instantaneous \ac{CSI} fluctuates at the wavelength scale, whereas these long-term statistics vary over much larger spatial and temporal scales.
        Thus, the optimization problem does not need to be solved at every channel coherence time, which significantly simplifies the implementation.
        Importantly, the proposed design is not a direct extension of conventional optimizations ~\cite{blockage_iimori,blockage_uchimura,blockagehybrid_uchimura,blockagewcnc_uchimura}; by exploiting long-term statistics to solve the discrete panel allocation as a combinatorial optimization problem, it achieves lightweight and robust beamforming design based on the theoretical analysis.

\end{itemize}

Through computational simulations, we evaluate the effectiveness of the theoretical analysis and the proposed optimal beamforming design with realistic millimeter-wave parameters.
The simulation results reveal that the conventional \ac{LoS}-concentrated beamforming design, while optimal for maximizing the average \ac{SE}, frequently causes outages of the target \ac{SE}, leading to unstable communications.
In contrast, the proposed multi-beam design, which utilizes both \ac{LoS} and \ac{NLoS} paths, can suppress the outage probability compared to the conventional method.

% Organization
The structure of this paper is outlined as follows. 
Section~\ref{sec:system} describes the system model, including the transmission frame, channel model and path blockage model, and the received signal, and Section~\ref{sec:beam} describes an overview of the analog beamforming design with multi-panel arrays.
In Section~\ref{sec:theory_SE}, the \ac{CDF} of the \ac{SE} is analytically derived for data transmission using a multi-panel array.  
The proposed beam design considering the outage probability and the average SE is detailed in Section~\ref{sec:panel_alloc}.  
In Section~\ref{sec:results}, the validity of the theoretical analysis is demonstrated, followed by clarifying that the proposed beam design leads to stable communication through computer simulations.  
Finally, the conclusions of this study are presented in Section~\ref{sec:conclusion}.

% Notation
\textit{Notation}: The following notations are used in this paper. 
Bold uppercase $\mathbf{X}$, bold lowercase $\mathbf{x}$, and non-bold lowercase $x$ are used to denote a matrix, a column vector, and a scalar value, respectively.
The imaginary unit is defined as $j = \sqrt{-1}$.
${(\cdot)}^\mathrm{T}$, ${(\cdot)}^\mathrm{H}$, ${(\cdot)}^\ast$, $|\cdot|$ and $\|\cdot \|_0$ denote the transpose, conjugate transpose, conjugate, absolute value, and $\ell_0$ norm, respectively.
$\mathbb{E}[\cdot]$ and $\circledast$ are operators of expectation and convolution.
$\mathbf{a}\circ \mathbf{b}$, and $\binom{a}{b}$ denote the Hadamard product and binomial coefficient between $\mathbf{a}$ and $\mathbf{b}$.
$f_X(x)$, $F_{X}(x)$ denote the \ac{PDF} and \ac{CDF} for the random variable $x$.
Especially, $\mathcal{CN}(x;\mu,v)$, $\mathcal{B}(x;p)$, and $\mathcal{U}(x;a,b)$ denote the circularly symmetric complex Gaussian distribution with mean $\mu$ and variance $v$, the Bernoulli distribution with probability $p$, and the uniform distribution between $a$ and $b$, respectively.
Moreover, $\mathcal{CN}\left(\mu,v\right)$, $\mathcal{B}(p)$ and $\mathcal{U}(a,b)$ represent random variables following the circularly symmetric complex Gaussian distribution with mean $\mu$ and variance $v$, the Bernoulli distribution with probability $p$, and the uniform distribution between $a$ and $b$.

\section{System Model}\label{sec:system}
Consider a point-to-point communication system between a single antenna \ac{UE} and a \ac{BS}.
The \ac{BS} is configured as a multi-panel array composed of $N_\mathrm{p}$ panels.
Each panel consists of a \ac{ULA} with $N_\mathrm{a}$ antenna elements and all panels are connected to a single \ac{RF} chain through power dividers.
As a result, the entire set of panels constitutes the \ac{ULA} with a total of $N_\mathrm{t} = N_\mathrm{a}N_\mathrm{p}$ elements.
The inter-panel and inter-element spacing is assumed to be set to half a wavelength, and phase calibration is assumed to be perfect, based on the prior works \cite{calibration_kabuto_TVT,calibration_kabuto,calibration_eberhardt,calibration_chen,mpCalibration_wang}.
Under these conditions, the \ac{BS} can independently generate multiple beams with each panel.
Moreover, as described above, assuming ideal phase calibration, the panels can be coordinated to align their directions to design a single sharp beam with one main lobe, or their directions can be independently controlled to design a multi-wide beam with multiple main lobes.

\par
This paper assumes a transmission frame consisting of two phases: the training phase and the transmission phase as shown in \figref{fig:frame_structure}.
In the training phase, beam sweeping is performed using predesigned candidate beams in a codebook to estimate the \ac{mmWave} channel.
Based on the estimated \ac{CSI}, the analog beam is designed for the subsequent transmission phase.
To consider training overhead in the beam sweeping, the analog beam remains fixed during the transmission phase without updates.
It is noted that the equivalent channel response observed in the digital domain can be continuously tracked by baseband signal processing using pilot signals inserted in the transmission phase.
Thus, in the transmission phase, a beam misalignment between the designed beam and the actual channel occurs due to the path blockages caused by the movement of obstacles such as pedestrians, vehicles, or falling leaves.
In the following subsections, we describe the \ac{mmWave} channel model in the training and transmission phases, and the received signal model in the transmission phase.

\begin{figure}[t]
  \begin{center}
      \includegraphics[width=\linewidth]{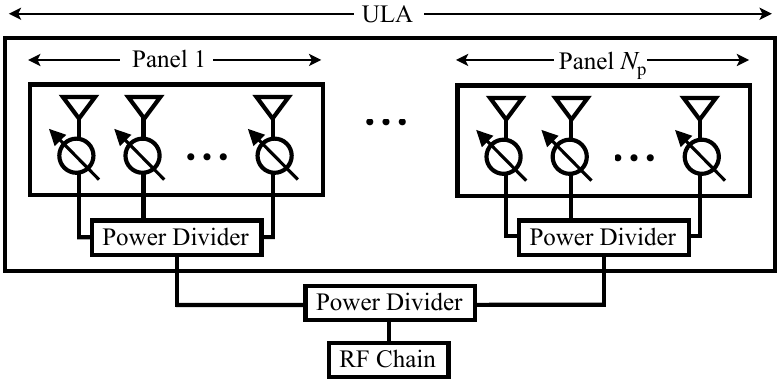}
  \end{center}
  \vspace*{-0.2cm}
  \caption{The BS structure with multi-panel array. Each panel, consisting of a ULA, is connected to a single RF chain through power dividers.}
  \label{fig:multipanel_structure}
\end{figure}

\subsection{Wireless Channel Modeling}
\subsubsection{Channel Model at the Training Phase}
\Ac{mmWave} channel at the training phase can be modeled as \cite{subarray_kamiwatari}
\begin{align}\label{eq:channel_training}
  \hat{\mathbf{h}} &= \sum_{l\in\mathcal{L}}g_l
  \,\mathbf{a}(N_\mathrm{t},~\theta_l)
  =\mathbf{A}({\boldsymbol{\theta}}){\mathbf{g}}
  \in\mathbb{C}^{N_\mathrm{t}},
\end{align}
where $L$, $l\in\mathcal{L}\triangleq\{1,2,\dots,L\}$ and $\theta_l\in[0,\pi)$ represent the number of paths, a path index and \ac{AoD} at the $l$-th path.
In the subsequent discussion, $l=1$ and $l\geq2$ are \ac{LoS} and \ac{NLoS} path index, respectively.
$\mathbf{a}(N,\theta)\triangleq
\begin{bmatrix}
1, \mathrm{e}^{j\pi\cos\theta}, \dots , \mathrm{e}^{j\pi(N-1)\cos\theta}
\end{bmatrix}^\mathrm{T}\in\mathbb{C}^{N}$
and 
$\mathbf{A}({\boldsymbol{\theta}})\triangleq
\begin{bmatrix}
\mathbf{a}(N_\mathrm{t}, {\theta}_1), \mathbf{a}(N_\mathrm{t}, {\theta}_2), \dots, \mathbf{a}(N_\mathrm{t}, {\theta}_L)
\end{bmatrix}\in\mathbb{C}^{N_\mathrm{t}\times L}$
represent array response vector and matrix
\footnote{
Wideband effects also exist.
However, since the main focus of this manuscript is to evaluate the impact of path blockage in millimeter-wave communications, the effect of beam squint is not considered.
}
, where $\boldsymbol{\theta}=\begin{bmatrix}\theta_1,\theta_2,\dots,\theta_L\end{bmatrix}^\mathrm{T}\in\mathbb{R}^{L}$
is an \ac{AoD} vector for all paths.
Here, $g_l$ denotes the path gain at the $l$-th path and it is modeled as Nakagami--Rice fading, treated as $g_l\sim \mathcal{CN}(0,\sigma^2_l)$, where $\sigma^2_l$ is a variance of the path gain for each path with Rician $K$-factor $\kappa$.
The variance $\sigma^2_l$ can be defined as
\begin{align}\label{eq:path_gain}
    \sigma_l^2 \triangleq \left\{
    \begin{aligned}
      &\dfrac{\kappa}{\kappa + 1},\quad l = 1~ (\mathrm{LoS})\\
      &\dfrac{1}{\kappa+1}\dfrac{1}{L-1},\quad l > 1~ (\mathrm{NLoS})\\
    \end{aligned}
    \right.~.
\end{align}
Let the path gain vector be denoted by $\mathbf{g}\triangleq\begin{bmatrix}g_1,g_2,\dots,g_L\end{bmatrix}\in\mathbb{C}^{L}$.
By exploiting the inherent angular sparsity in \ac{mmWave} bands, the channel vector in \eqref{eq:channel_training} can be efficiently estimated using \ac{CS} techniques with training pilots during the beam sweeping process~\cite{overview_heath,channel_estimation_rodriguez,channel_estimation_andrei}.
Since the \ac{CS}-based approaches can achieve highly accurate channel estimation typically yielding a \ac{NMSE} of around –20 dB and the main focus of this paper is to establish a theoretical foundation for multi-beam design, we assume that channel estimation in the training phase is ideal without estimation errors, following prior studies~\cite{blockage_iimori,blockage_uchimura,blockagewcnc_uchimura,blockagehybrid_uchimura}.

\subsubsection{Channel Model at the Transmission Phase}
In the subsequent transmission phase, obstructions may block the paths abruptly\cite{rapid_maccartney,empirical_slezak,humanblockage_mukherjee}.
Channel response vector with stochastic path blockage can be modeled as
\begin{align}\label{eq:channel_transmission}
  \mathbf{h} &= \sum_{l\in\mathcal{L}}\omega_lg_l
  \,\mathbf{a}(N_\mathrm{t},~\theta_l)
  =\mathbf{A}(\boldsymbol{\theta})(\boldsymbol{\omega}\circ\mathbf{g})
  \in\mathbb{C}^{N_\mathrm{t}},
\end{align}
where $\omega_l$ denotes the blockage parameter for the $l$-th path and is independently defined across paths, as in \cite{blockage_uchimura}:
\begin{subequations}\label{eq:blockage_real}
    \begin{align}\label{eq:omega_blocked}
        &\mathrm{Pr} \{ \omega_l = 1/\eta \} = \hat{p}_l \quad \quad (\mathrm{Blocked}),\\
        \label{eq:omega_otherwise}
        &\mathrm{Pr} \{ \omega_l = 1 \} = 1 - \hat{p}_l \quad \ (\mathrm{Otherwise}),
    \end{align}
\end{subequations}
where $\eta\in\mathbb{R}$ and $\hat{p}_l\in\mathbb{R}$ denote the blockage attenuation and blockage probability for the $l$-th path. 
{In this paper, blockage events are assumed to be independent across propagation paths for analytical tractability.
In practical environments, however, common or moving obstacles may induce spatially and temporally correlated blockage and channel variations~\cite{3GPP_LSP,blockage_temporal}.
The present model focuses on the fundamental impact of stochastic path blockage on beamforming without specifying site- and mobility-dependent correlations.
Let $\boldsymbol{\omega} \triangleq \begin{bmatrix}\omega_1,\omega_2,\dots,\omega_L\end{bmatrix}^\mathrm{T}\in\mathbb{R}^{L}$ denote the vector of blockage parameters.
Since the blockage effects depend on the beam width~\cite{pathloss_rappaport,pblk_raghavan, rapid_maccartney}, the blockage attenuation $\eta$ is given by $\eta = 9.8 + {180^\circ} / A_\mathrm{BW}$ with \ac{HPBW} of analog beamforming $A_\mathrm{BW}~[\mathrm{deg}]$~\cite{rapid_maccartney}.

\subsection{Received Signal During the Transmission Phase}
In the transmission phase, the received signal $y$ is given as 
\begin{align}\label{eq:received_signal}
    y = \mathbf{h}^\mathrm{H}\mathbf{f}s + z = h_\mathrm{eq}s + z \in\mathbb{C},
\end{align}
where $\mathbf{f}\in\mathbb{C}^{N_\mathrm{t}}$ is an analog beamforming vector designed by estimated CSI, $s\in\mathcal{CN}(0, P_\mathrm{tx})$ is a transmit signal with average transmit power $P_\mathrm{tx}$, and $z\in\mathcal{CN}(0,\sigma_z^2)$ is an \ac{AWGN} with noise variance $\sigma_z^2$, respectively.
$h_\mathrm{eq} \in \mathbb{C}$ is an equivalent channel response including the analog beamforming, which can be written as
\begin{align}\label{eq:equivalent_channel}
    h_\mathrm{eq} \triangleq \mathbf{h}^\mathrm{H}\mathbf{f} =(\boldsymbol{\omega}\circ\mathbf{g})^\mathrm{H} \mathbf{a}_\mathrm{eq},
\end{align}
where $\mathbf{a}_\mathrm{eq} \triangleq \mathbf{A}^\mathrm{H}(\boldsymbol{\theta})\mathbf{f}\in\mathbb{C}^{L}$ is an equivalent array response.
\Ac{SNR} is defined as $\gamma_\mathrm{tx}\triangleq P_\mathrm{tx}/\sigma_z^2$.
In the following section, it is assumed that the equivalent channel response $h_\mathrm{eq}$ can be ideally estimated using baseband signal processing with pilot signals inserted in the transmission phase\footnote{
Note that the number of pilots required to estimate $h_\mathrm{eq}$ is significantly smaller than that needed for beam sweeping because the equivalent channel response $h_\mathrm{eq}$ is a scalar quantity.
Thus, the pilot overhead for baseband channel estimation is negligible.
}.

\begin{figure}[t]
  \begin{center}
      \includegraphics[width=\linewidth]{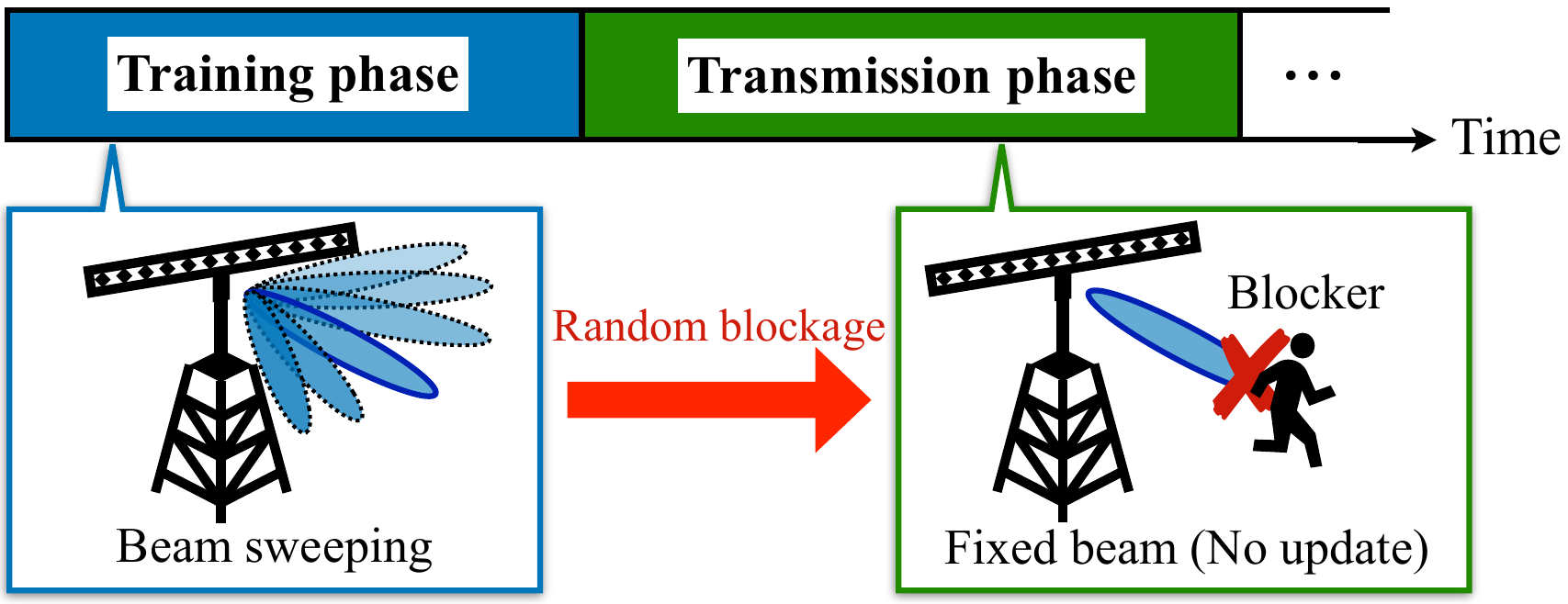}
  \end{center}
  \vspace*{-0.2cm}
  \caption{Illustration of the transmission frame structure and beamforming design. If unexpected path blockage occurs after the training phase, a beam misalignment arises between the designed beam and the propagation channel during the transmission phase, which degrades the communication quality.}
  \label{fig:frame_structure}
\end{figure}

\section{Analog Beamforming Design}\label{sec:beam}
In this section, we introduce the analog beamforming design with multi-panel arrays.
By leveraging the structure of the multi-panel array, each panel can independently generate the analog beamforming.
Using the estimated \ac{CSI} at the training phase, the analog beamforming vector $\mathbf{f}_m\in\mathbb{C}^{N_\mathrm{a}}$ at the $m$-th panel can be designed as
\begin{align}\label{eq:beamforming_panel}
    \mathbf{f}_\mathrm{m} = \dfrac{\psi_m}{\sqrt{N_\mathrm{t}}}\mathbf{a}(N_\mathrm{a},\phi_m)\in\mathbb{C}^{N_\mathrm{a}},
\end{align}
where $\psi_m$ and $\phi_m$ denote a phase difference and directivity at the $m$-th panel, respectively.
The phase difference at the $m$-th panel is defined as
\begin{align}\label{eq:phase_difference}
    \psi_m = \mathrm{exp}\{j\pi(m-1)N_\mathrm{a}\cos{(\phi_m)}\}.
\end{align}

Furthermore, we assume that the \ac{AoD} of each path $\theta_l$ can be estimated perfectly with high precision beam sweeping in the training phase.
Hence, the \ac{BS} can generate ideal analog beams aligned with the \ac{AoD}.
By combining the beamforming vectors for each panel as described in \eqref{eq:beamforming_panel}, the analog beamforming vector $\mathbf{f}$ can be represented as
\begin{align}\label{eq:beamforming_all}
    \mathbf{f} = [\mathbf{f}_1^\mathrm{T},\dots,\mathbf{f}_{N_\mathrm{p}}^\mathrm{T}]^\mathrm{T}\in\mathbb{C}^{N_\mathrm{t}}.
\end{align}

Hereafter, the proposed analog beamforming design is discussed in detail in comparison with conventional methods.
Conventional beamforming designs align all panels cooperatively to form a sharp beam that focuses exclusively on the \ac{LoS} component of the \ac{AoD}, $\theta_1$ such that $\phi_m = \theta_1,\ \forall m \in \{1,\ldots, N_\mathrm{p} \}$ in \eqref{eq:beamforming_panel}.
In contrast, the proposed method leverages the estimated \acp{AoD} of all paths, $\boldsymbol{\theta}$, obtained during the training phase, to enable the design of multiple beams that align with multiple \acp{AoD}.
\figref{fig:beampattern_exact} illustrates the beam patterns of the conventional analog beam with a single directionality and the proposed analog beam with multiple directionalities enabled by independent panel control.
As shown in the figure, the conventional beam focuses only on the LoS direction with high array gain, leading to frequent communication disconnections if the LoS path is blocked by obstacles.
In contrast, the proposed method gains spatial diversity owing to multiple beams aligned with multiple paths, enabling stable communication through the remaining paths even if any path is blocked.
However, owing to the trade-off between array gain and spatial diversity, determining the number of panels allocated to each path is essential to improving system performance.
To represent the number of panels assigned to each path, the panel allocation vector $\mathbf{q} = {\begin{bmatrix}q_1, q_2, \dots, q_L
\end{bmatrix}}^\mathrm{T}\in\mathbb{Z}^L$ is introduced.
$q_l \geq 0$ is the number of panels allocated to the $l$-th path such that $\sum_{l\in\mathcal{L}} q_l = N_\mathrm{p}$.
For the sake of future convenience, let $N_\mathrm{b}$ denote the number of beams aligned with the selected $N_\mathrm{b}$ paths, where $N_\mathrm{b}$ satisfies $N_\mathrm{b}=\|\mathbf{q}\|_0$.
It should be noted that this paper focuses on far-field communications, where the signal wavefront can be approximated by a plane wave.
Under this assumption, the \acp{AoD} are identical across all panels.
Consequently, the beam gain for a given path depends solely on the number of panels allocated to that path, and does not depend on the physical positions of the allocated panels.

The panel allocation vector $\mathbf{q}$ is not uniquely determined and must be carefully designed to align with the specific objectives of the system.
Therefore, we optimize panel allocation to minimize the outage probability of the SE, which is a metric of communication stability.
To formulate the objective function for the outage probability, the distribution of SE should be derived in closed form, which is described in detail in Section~\ref{sec:theory_SE}.
The optimization method for panel allocation using the derived outage probability is described in Section~\ref{sec:panel_alloc}.

\section{Theoretical Analysis of Spectral Efficiency}
\label{sec:theory_SE}
In this section, we derive the CDF of the \ac{SE}, when using the multi-beam design with multimodal directivity, in the considered mmWave communication system with the multi-panel array.
Based on the received signal model in~\eqref{eq:received_signal}, the \ac{SE} $\xi$ can be expressed as
\begin{align}
    \label{eq:SE_difinition}
    \xi    &\triangleq \log_2(1 + \gamma),
\end{align}
where $\gamma$ denotes the \ac{RSNR}, which can be expressed as 
\begin{align}\label{eq:totalSNR}
    \gamma &\triangleq \gamma_\mathrm{tx}|h_\mathrm{eq}|^2 = \gamma_\mathrm{tx}|(\boldsymbol{\omega} \circ \mathbf{g})^\mathrm{H}\mathbf{a}_\mathrm{eq}|^2.
\end{align}

To derive the CDF of the SE, the distribution of the equivalent channel $h_\mathrm{eq}$ must be derived.
By applying an approximation~\cite{sinc_wang} for array responses, the distribution of $h_\mathrm{eq}$ can be approximately derived, as detailed in Section~\ref{subsec:approx_array}.
From the derived distribution of $h_\mathrm{eq}$, the CDF of the SE can be derived, detailed in Section~\ref{subsec:derivation_RSNR}.

\begin{figure*}[t]
    \begin{minipage}[b]{0.48\textwidth}
        \begin{center}
            \includegraphics[width=\linewidth]{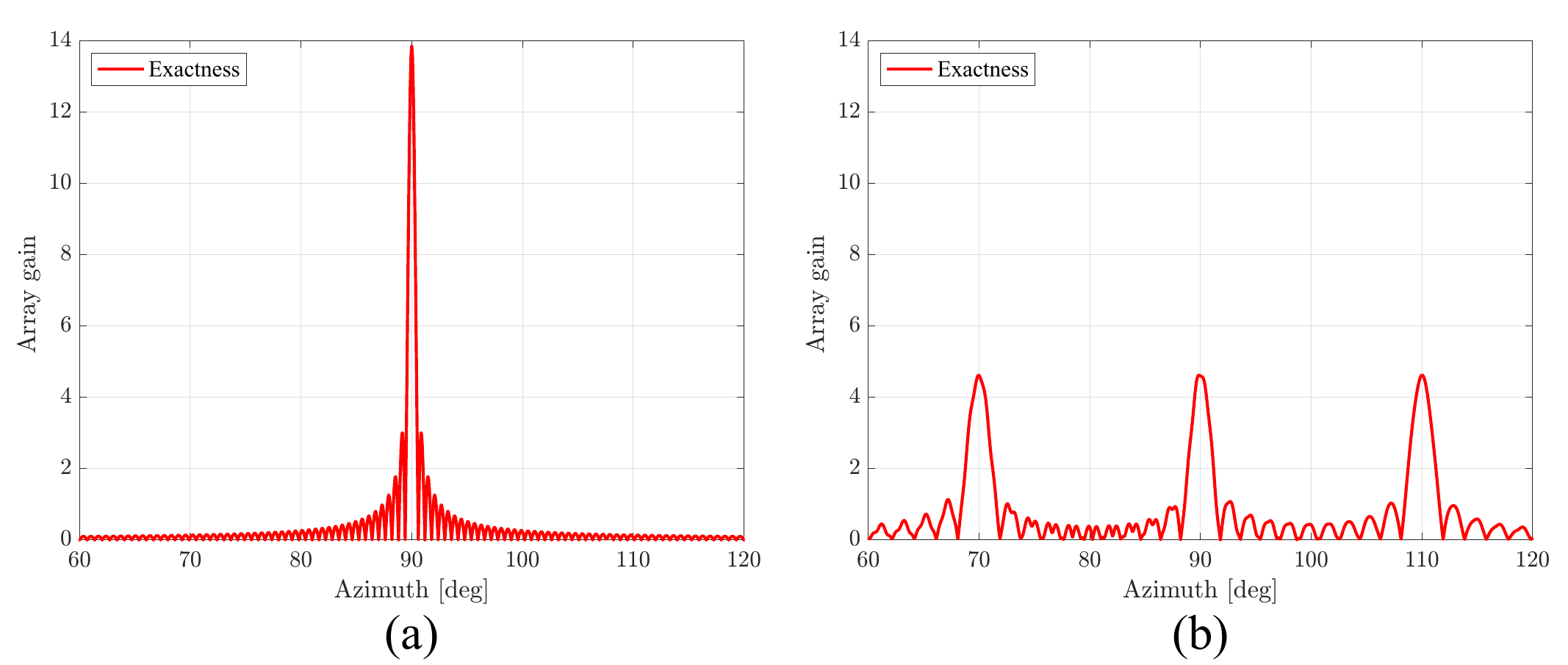}
        \end{center}
        \vspace*{-0.2cm}
        \caption{Comparison of the beam pattern between (a) the conventional single main lobe beamforming with $\mathbf{q} = [N_\mathrm{p}, 0, 0]^\mathrm{T}$, and (b) the proposed multi-beam with $\mathbf{q} = [2, 2, 2]^\mathrm{T}$, where $N_\mathrm{a} = 32$, $N_\mathrm{p} = 6$.}
      \label{fig:beampattern_exact}
    \end{minipage}
    \hspace{5mm}
    \begin{minipage}[b]{0.48\textwidth}
        \begin{center}
            \includegraphics[width=\linewidth]{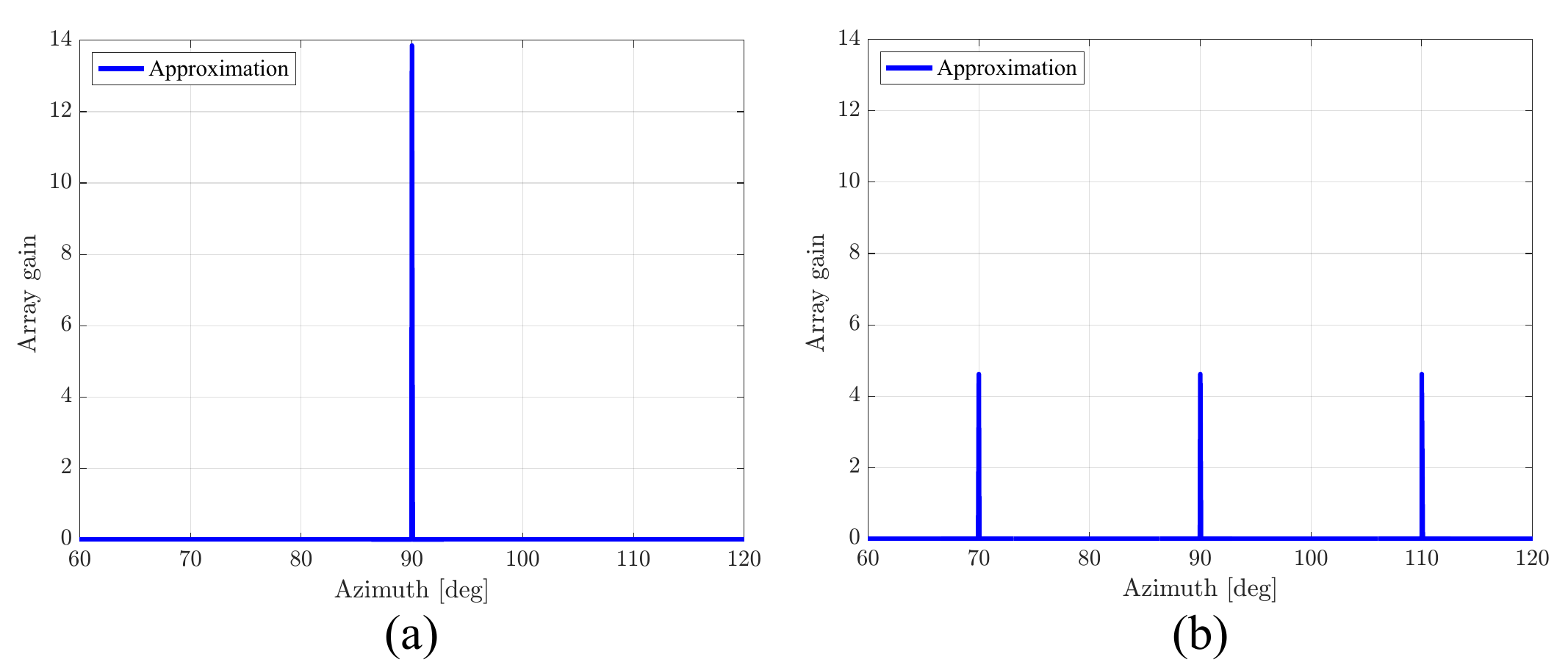}
        \end{center}
        \vspace*{-0.2cm}
        \caption{Comparison of the approximated beam pattern between (a) the conventional beam with $\mathbf{q} = [N_\mathrm{p}, 0, 0]^\mathrm{T}$, and (b) the proposed multi-beam with $\mathbf{q} = [2, 2, 2]^\mathrm{T}$, where $N_\mathrm{a} = 32$, $N_\mathrm{p} = 6$.}
        \label{fig:beampattern_approx}
    \end{minipage}
    % \caption{CDF of the SE for Multiple Target SE}
\end{figure*}

\par
\subsection{Approximation of the Blockage Parameters}
\label{subsec:approx_omega}
Path blockages arise stochastically as a result of temporal fluctuation in the communication environment around the \ac{BS} and \ac{UE}.
Under the pessimistic assumption that the energy completely vanishes without diffraction when a path is blocked, the blockage attenuation is set to $\eta \to \infty$.
In this case, blockage parameter $\omega_l$ can be approximated as binary values $\tilde{\omega}_l\in\{0,1\}$ and follows a Bernoulli distribution with blockage probability $\hat{p}_l$~\cite{blockage_iimori,blockagewcnc_uchimura,blockage_uchimura,blockagehybrid_uchimura}.
The \ac{PDF} of the $\hat{p}_l$ can be expressed as $f_{\hat{P}_l}(\hat{p}_l) = \mathcal{U}(\hat{p}_l;p_\mathrm{min}, p_\mathrm{max}),~\forall l$, where $p_\mathrm{min}$ and $p_\mathrm{max}$ are minimum and maximum blockage probability, respectively.
The \acp{PMF} of $\tilde{\omega}_l$ and $\tilde{\boldsymbol{\omega}} \triangleq \begin{bmatrix}\tilde{\omega}_1,\tilde{\omega}_2,\dots,\tilde{\omega}_L\end{bmatrix}^\mathrm{T}\in\mathbb{R}^{L}$, given the blockage probability $\hat{p}_l$, are expressed as
\begin{subequations}
    \label{eq:blockage_PMF}
    \begin{align}
    \label{eq:blockage_PMF_path}
    f_{\tilde{\Omega}_l}(\tilde{\omega}_l|\hat{p}_l) &= \hat{p}_l^{(1-\tilde{\omega}_l)}(1-\hat{p}_l)^{\tilde{\omega}_l}, \tilde{\omega}_l\in\{0,1\}, \forall l, \\
    \label{eq:blockage_PMF_vector}
    f_{\tilde{\Omega}}(\tilde{\boldsymbol{\omega}}|\hat{p}_l) &= \prod_{l=1}^L f_{\tilde{\Omega}_l}(\tilde{\omega}_l|\hat{p}_l).
    \end{align}
\end{subequations}

By marginalizing \eqref{eq:blockage_PMF} over the blockage probability $\hat{p}_l$, the \acp{PMF} of the blockage parameters can be reformulated as
\begin{subequations}
    \label{eq:blockage_PMF_marge}
    \begin{align}
    \label{eq:blockage_PMF_marge_path}
    f_{\tilde{\Omega}_l}(\tilde{\omega}_l) &= \int f_{\tilde{\Omega}_l}(\tilde{\omega}_l|\hat{p}_l) f_{\hat{P}_l}(\hat{p}_l) d\hat{p}_l, \nonumber \\
    &= p_\mathrm{blk}^{(1-\tilde{\omega}_l)}(1-p_\mathrm{blk})^{\tilde{\omega}_l}
    = \mathcal{B}(\tilde{\omega}_l ; p_\mathrm{blk}), \\
    \label{eq:blockage_PMF_marge_vector}
    f_{\tilde{\Omega}}(\tilde{\boldsymbol{\omega}}) &= \prod_{l=1}^L f_{\tilde{\Omega}_l}(\tilde{\omega}_l),
    \end{align}
\end{subequations}
where $p_\mathrm{blk}$ is an average path blockage probability, given by $p_\mathrm{blk} = (p_\mathrm{min} + p_\mathrm{max})/2$.

\par
\subsection{Approximation of the Equivalent Array Response}
\label{subsec:approx_array}
In~\eqref{eq:totalSNR}, the equivalent array response $\mathbf{a}_\mathrm{eq} = \mathbf{A}^\mathrm{H}(\boldsymbol{\theta}) \mathbf{f}$ is composed of the inner products between the array responses and the analog beamforming vector, which is also the array response, as shown in~\eqref{eq:beamforming_panel}.
To simplify the expression of the inner products for array responses, this paper adopts the following approximation~\cite{sinc_wang}, which is valid in large-scale array systems.
\begin{align}\label{eq:array_response_approx}
    \mathbf{a}(N_\mathrm{a},~\theta)^\mathrm{H}\mathbf{a}(N_\mathrm{a},\phi)\approx
    \left\{
        \begin{aligned}
        &N_\mathrm{a},~~(\theta=\phi,~\mathrm{main~lobe})\\
        &0,\  \quad (\theta\neq\phi,~\mathrm{side~lobe})\\
        \end{aligned}
    \right. .
\end{align}

\par
This approximation implies that the main lobe contains the effective energy, while the remaining parts, including the side lobes, are assumed to be zero as shown in \figref{fig:beampattern_approx}.
Using the approximation in \eqref{eq:array_response_approx}, the equivalent array response $\mathbf{a}_\mathrm{eq}$ can be approximated with panel allocation vector $\mathbf{q}$ as
\begin{align}\label{eq:equivalent_array_response_approx}
    \mathbf{a}_\mathrm{eq} &=
    \begin{bmatrix}
    \mathbf{a}(N_\mathrm{t},\theta_1),\dots,\mathbf{a}(N_\mathrm{t},\theta_L)
    \end{bmatrix}^\mathrm{H} \nonumber\\
    &\quad \cdot~
    \dfrac{1}{\sqrt{N_\mathrm{t}}}
    \begin{bmatrix}
    \psi_1\mathbf{a}(N_\mathrm{a},\phi_1)^\mathrm{T},\dots,\psi_{N_\mathrm{p}}\mathbf{a}(N_\mathrm{a},\phi_{N_\mathrm{p}})^\mathrm{T}
    \end{bmatrix}^\mathrm{T},\nonumber\\
    &\approx
    \dfrac{N_\mathrm{a}}{\sqrt{N_\mathrm{t}}}
    \mathbf{q}.
\end{align}

As evaluated in the numerical simulation described in Section~\ref{sec:results}, the effects of the approximation error on the CDF of the SE are negligible.

\subsection{Derivation of the Equivalent Channel Response and RSNR}
\label{subsec:derivation_RSNR}
Substituting the equivalent array response in \eqref{eq:equivalent_array_response_approx} into \eqref{eq:equivalent_channel},
the  equivalent channel response $h_\mathrm{eq}$ with binary blockage approximation $\tilde{\omega}_l$ can be approximated as
\begin{align}\label{eq:heq_approx}
    h_\mathrm{eq}&\approx
    \dfrac{N_\mathrm{a}}{\sqrt{N_\mathrm{t}}}\sum_{l\in\mathcal{L}} g_l^\ast \tilde{\omega}_l q_l
    = \dfrac{N_\mathrm{a}}{\sqrt{N_\mathrm{t}}}\sum_{l\in\mathcal{L}} g_{q_l} \tilde{\omega}_l ,
\end{align}
where $g_{q_l}$ is a random variable and can be represented as $g_{q_l} \sim \mathcal{CN}(0, \rho_l^2)$ with variance $\rho_l^2 \triangleq \sigma^2_l q_l^2$.
As shown in equation \eqref{eq:heq_approx}, the random variable of the equivalent channel response $h_\mathrm{eq}$ is expressed as the sum of the products of the Gaussian random variable $g_{q_l}$ and the Bernoulli random variable $\tilde{\omega}_l$.
Therefore, the \ac{PDF} of $h_\mathrm{eq}$ becomes the Bernoulli--Gaussian mixture, which is expressed as \eqref{eq:PDF_heq} in the top of next page.
The detail of the derivation of \eqref{eq:PDF_heq} is described in \appref{app:heq}.
\figref{fig:PDF_equivalent_channel} illustrates the \ac{PDF} of the equivalent channel response, where the number of paths is $L=2$ with the LoS path and one NLoS path. 
Because of $L=2$, there are $2^L=4$ blockage patterns: 1) both paths are blocked, 2) only the NLoS path is blocked, 3) only the LoS path is blocked, and 4) neither path is blocked.
Subsequently, the PDF of the equivalent channel $f_{H_\mathrm{eq}}(h_\mathrm{eq})$ is expressed as the summation of four distributions $f_1,\dots,f_4$ corresponding to each blockage pattern, as shown in \figref{fig:PDF_equivalent_channel}.
\begin{figure}[t]
    \begin{center}
      \includegraphics[width=\linewidth]{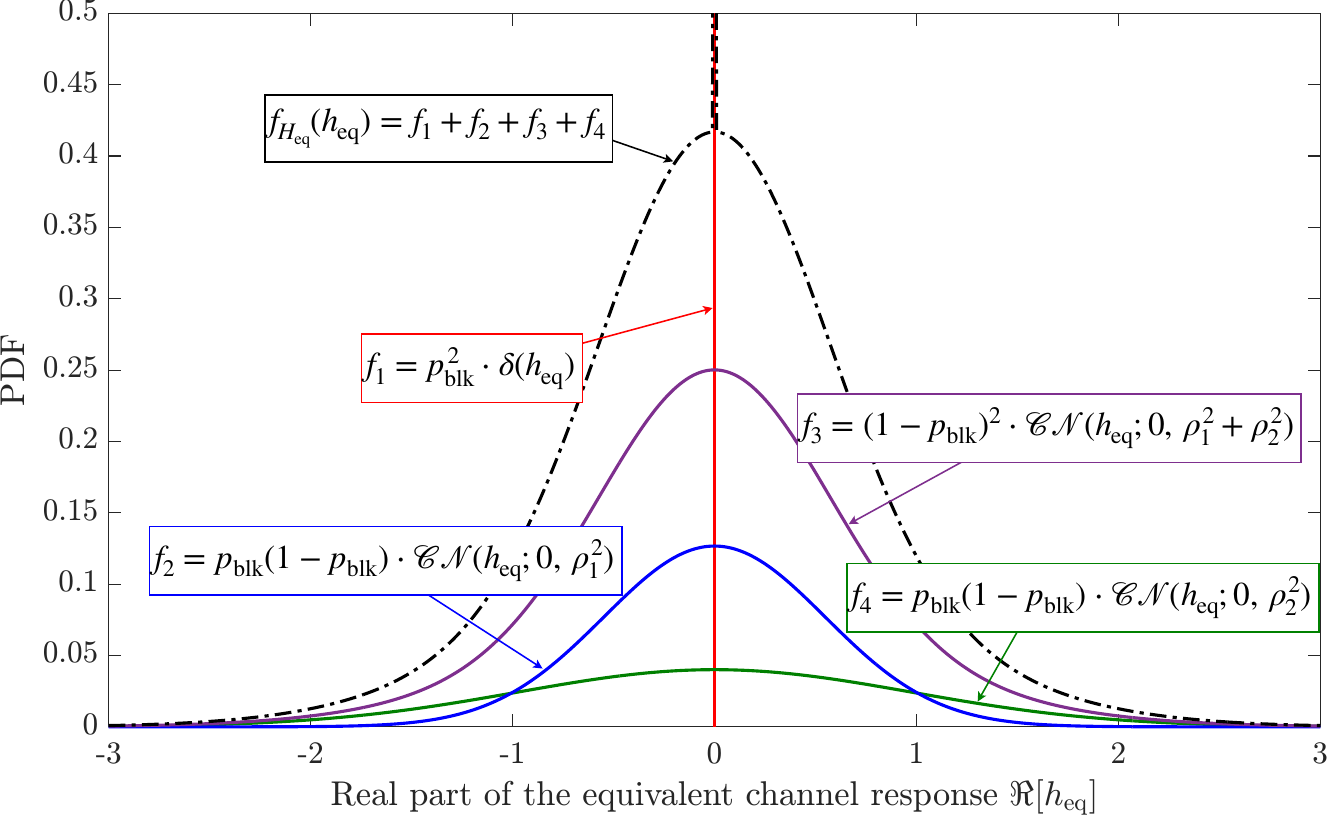}
  \end{center}
    \vspace*{-0.2cm}
    \caption{PDF of the real part of the equivalent channel response $\Re[h_\mathrm{eq}]$ with  $L=2$, $N_\mathrm{a}=32$, $N_\mathrm{p} = 4$, $p_\mathrm{blk} = 0.4$, $\kappa = 10~[\mathrm{dB}]$, $\mathbf{q} = [2, 2]^\mathrm{T}$, $N_\mathrm{b} = 2$}
    \label{fig:PDF_equivalent_channel}
\end{figure}

\begin{figure*}[t]
    \begin{align}\label{eq:PDF_heq}
        f_{H_\mathrm{eq}}(h_\mathrm{eq}) = p_\mathrm{blk}^{N_\mathrm{b}} \delta (h_\mathrm{eq}) + \sum_{t=1}^{N_\mathrm{b}} p_\mathrm{blk}^{N_\mathrm{b}-t}(1-p_\mathrm{blk})^t 
        \left\{
            \sum_{1\leq i_1<\dots<i_t\leq L} \mathcal{CN}\left(h_\mathrm{eq}; 0, \dfrac{N_\mathrm{a}^2}{N_\mathrm{t}}\sum_{k=1}^t\rho_{i_k}^2\right)
        \right\}
    \end{align}
    %\rule{\textwidth}{0.4pt} 
    \vspace{-2em} 
\end{figure*}

With an approximation of the equivalent channel response $h_\mathrm{eq}$ in \eqref{eq:heq_approx}, the \ac{RSNR} $\gamma$ defined in \eqref{eq:totalSNR} can also be approximated as
\begin{align}\label{eq:totalSNR_approx} 
\gamma=
\gamma_\mathrm{tx} |h_\mathrm{eq}|^2
\approx \dfrac{\gamma_\mathrm{tx} N_\mathrm{a}^2}{N_\mathrm{t}}\left|\sum_{l\in\mathcal{L}} g_{q_l} \tilde{\omega}_l \right|^2.
\end{align}

As shown in \eqref{eq:totalSNR_approx}, the distribution of the \ac{RSNR} $\gamma$ can be represented as the square of a Bernoulli--Gaussian mixture random variable.
The \ac{PDF} and \ac{CDF} of the \ac{RSNR} $\gamma$ are derived in \eqref{eq:PDF_totalSNR} and \eqref{eq:CDF_totalSNR}, respectively. 
The detailed derivations of \eqref{eq:PDF_totalSNR} and \eqref{eq:CDF_totalSNR} are provided in \appref{app:RSNR}.
Subsequently, by substituting $\gamma = 2^\xi -1$ into \eqref{eq:CDF_totalSNR}, the \ac{CDF} of the \ac{SE} can be derived in closed form because the logarithm function is a monotonically increasing function.

\begin{figure*}[t]
    \begin{align}\label{eq:PDF_totalSNR}
    % \Kanta{
        f_{\Gamma}(\gamma) = p_\mathrm{blk}^{N_\mathrm{b}} \delta (\gamma) + \sum_{t=1}^{N_\mathrm{b}} p_\mathrm{blk}^{N_\mathrm{b}-t}(1-p_\mathrm{blk})^t 
        \left\{
            \sum_{1\leq i_1<\dots<i_t\leq L}
            \dfrac{1}{\gamma_\mathrm{tx}N_\mathrm{a}^2/N_\mathrm{t}\sum_{k=1}^t\rho_{i_k}^2~}~ \mathrm{exp}\left(- \dfrac{\gamma}{\gamma_\mathrm{tx}N_\mathrm{a}^2/N_\mathrm{t}\sum_{k=1}^t\rho_{i_k}^2}\right)
        \right\}
    % }
    \end{align}
    %\rule{\textwidth}{0.4pt} 
    \vspace{-1em} 
\end{figure*}

\begin{figure*}[t]
\begin{align}\label{eq:CDF_totalSNR}
    F_\Gamma(\gamma) = \int_{-\infty}^{\gamma} f_{\Gamma}(\tilde{\gamma}) d \tilde{\gamma} = 
    p_\mathrm{blk}^{N_\mathrm{b}} + \sum_{t=1}^{N_\mathrm{b}} p_\mathrm{blk}^{N_\mathrm{b}-t}(1-p_\mathrm{blk})^t \left\{ \sum_{1\leq i_1<\dots<i_t\leq L} 
    1- \mathrm{exp}\left(-\dfrac{\gamma}{\gamma_\mathrm{tx}N_\mathrm{a}^2/N_\mathrm{t}\sum_{k=1}^{t}\rho_{i_k}^2}\right) \right\}
\end{align}
% \rule{\textwidth}{0.4pt} 
\vspace{-2em} 
\end{figure*}

\section{Panel Allocation Method}
\label{sec:panel_alloc}
In this section, we consider three analog beamforming designs using the distributions of the SE derived in Section~\ref{subsec:derivation_RSNR}. 
The first method considers only the maximization of the average SE, which is confirmed to be equivalent to the conventional beamforming design focusing only on the LoS path, as described in Section~\ref{subsec:average_SE}.
The second method considers only the minimization of the outage probability, which is described in Section~\ref{subsec:outage_min}.
The third method considers both the average SE and the outage probability, as described in Section~\ref{subsec:outmin_ASE}.

\subsection{Average SE Maximization}
\label{subsec:average_SE}
This subsection formulates the panel allocation problem based on maximizing the average \ac{SE}.
The average \ac{SE} is defined as $\mathbb{E}[\xi] = \mathbb{E}\left[\log_2 (1 + \gamma)\right]$.
Using Jensen's inequality \cite{jensen} because of the strict concavity of the logarithmic function, the average \ac{SE} is bounded as
\begin{align}\label{eq:average_SE_Jensen}
    \mathbb{E}[\xi] &= \mathbb{E}\left[\log_2 (1 + \gamma)\right]
    \leq \log_2 \left(1 + \mathbb{E}[\gamma]\right) = \xi_\mathrm{ave},
\end{align}
where $\xi_\mathrm{ave}$ is the upper bound of the average \ac{SE} corresponding to the SE at the average \ac{RSNR} $\mathbb{E}[\gamma]$.

For analytical tractability, we maximize the upper bound of the SE $\xi_\mathrm{ave}$ instead of directly maximizing the average SE $\mathbb{E}[\xi]$.
From \eqref{eq:average_SE_Jensen}, the maximization of the upper bound $\xi_\mathrm{ave}$ is equivalent to the maximization of the average \ac{RSNR} $\mathbb{E}[\gamma]$.
Subsequently, the panel allocation problem to maximize the upper bound is formulated as
\begin{subequations}\label{opt:average_SNR}
    \begin{align}
        \underset{\mathbf{q}\in\mathbb{Z}^L}{\mathrm{maximize}} &\quad \mathbb{E}[\gamma],\\
        \mathrm{subject~to}& \quad q_l \geq 0,\\
        &\quad \sum_{l\in\mathcal{L}} q_l = N_\mathrm{p}.
    \end{align}
\end{subequations}

From \eqref{eq:PDF_heq}, the average \ac{RSNR} given the random variables $\tilde{\boldsymbol{\omega}}$, denoted as $\mathbb{E}[\gamma | \tilde{\boldsymbol{\omega}}]$, can be expressed as
\begin{align}\label{eq:totalSNR_expectation_given_omega}
    \mathbb{E}[\gamma | \tilde{\boldsymbol{\omega}}] = \gamma_\mathrm{tx} \mathbb{E}\left[ \left| h_{\mathrm{eq}}\right|^2 \Big | \tilde{\boldsymbol{\omega}} \right]= \dfrac{\gamma_\mathrm{tx}N_\mathrm{a}^2}{N_\mathrm{t}}\sum_{l\in\mathcal{L}}\tilde{\omega_l}^2 q_l^2 \sigma_l^2.
\end{align}

\par
As each path blockage parameter $\tilde{\omega}_l$ is discrete random variable as defined in \eqref{eq:blockage_PMF}, the average \ac{RSNR} $\mathbb{E}[\gamma]$ can be calculated by averaging $\mathbb{E}[\gamma | \tilde{\boldsymbol{\omega}}]$ over the random variable $\tilde{\boldsymbol{\omega}}$ as \eqref{eq:totalSNR_expectation} in the top of next page, where $t$ denotes the number of non-blocked paths.
\begin{figure*}[t]
\begin{align}\label{eq:totalSNR_expectation}
    \mathbb{E}[\gamma] &= 
    \sum_{\tilde{\boldsymbol{\omega}} \in \{0,1\}^{L \times 1} } \mathbb{E}[\gamma | \tilde{\boldsymbol{\omega}}] p(\tilde{\boldsymbol{\omega}})
    = 
    \dfrac{\gamma_\mathrm{tx}N_\mathrm{a}^2}{N_\mathrm{t}} \sum_{t=1}^L p_\mathrm{blk}^{L-t} (1 - p_\mathrm{blk})^t\sum_{1\leq i_1 < \dots < i_t\leq L } \sum_{k=1}^t  q_{i_k}^2 \sigma_{i_k}^2\nonumber \\
    &= \dfrac{\gamma_\mathrm{tx}N_\mathrm{a}^2}{N_\mathrm{t}(\kappa+1)(L-1)} \left(1 - p_\mathrm{blk}\right)\left(\kappa(L-1) q_1^2 + q_2^2 + \dots + q_L^2\right)
\end{align}
\rule{\textwidth}{0.4pt} 
% \vspace{1em} 
\end{figure*}

\par
From the average RSNR $\mathbb{E}[\gamma]$ in \eqref{eq:totalSNR_expectation}, the maximization problem in \eqref{opt:average_SNR} can be reformulated as

\begin{subequations}\label{opt:average_SNR_dominant}
    \begin{align}
        \label{eq:obj_avgSE}
        \underset{\mathbf{q}\in\mathbb{Z}^L}{\mathrm{maximize}} &\quad \kappa(L-1) q_1^2 + q_2^2 + \dots + q_L^2,\\
        \mathrm{subject~to}& \quad q_l \geq 0,\\
        &\quad \sum_{l\in\mathcal{L}} q_l = N_\mathrm{p}.
    \end{align}
\end{subequations}

The objective function of \eqref{opt:average_SNR_dominant} is expressed as a sum of squared allocation variables $q_l$.
As the Rician $K$-factor $\kappa$ is typically larger than 1 in practical \ac{mmWave} communication environments\cite{kfactor_samimi}, 
the first term of the objective function in \eqref{eq:obj_avgSE}, $\kappa(L-1) q_1^2$ is dominant for the maximization problem.
Thus, the optimal solution of the maximization problem \eqref{opt:average_SNR_dominant} can be derived in closed form as
\begin{align}\label{eq:average_SE_solusion}
    \mathbf{q} = \begin{bmatrix}
        N_\mathrm{p},~ 0,~ \dots,~ 0
    \end{bmatrix}^\mathrm{T}.
\end{align}

This solution indicates that focusing all panels only on the LoS path can maximize the average \ac{RSNR}, which corresponds to maximizing the upper bound of the average SE.
Therefore, the codebook-based beamforming used in current \ac{mmWave} systems is optimal for maximizing the average \ac{RSNR}.

\subsection{Outage Probability Minimization}
\label{subsec:outage_min}
To achieve stable communication, it is essential not only to enhance the average \ac{SE} but also to mitigate the tail of its distribution, corresponding to the occurrence probability of poor SE performance.
Therefore, we propose the panel allocation that minimizes the outage probability for a desired target \ac{SE}.
In this paper, the outage probability is defined as
\begin{align}\label{eq:outage_probability}
    p_\mathrm{out} \triangleq \mathrm{Pr}\{\log_2(1 + \gamma) \leq \xi_\mathrm{th}\},
\end{align}
where $\xi_\mathrm{th}$ denotes the target \ac{SE} desired by the \ac{UE}.
Using \eqref{eq:outage_probability}, the panel allocation problem for minimizing the outage probability can be formulated as
\begin{subequations}\label{opt:outage_SE_minimization}
    \begin{align}
        \underset{q_l,~\forall l\in\mathcal{L}}{\mathrm{minimize}}
        \quad& \mathrm{Pr}\{\log_2(1 + \gamma) \leq \xi_\mathrm{th}\},\\
        \mathrm{subject\:to}
        \quad
        & \sum_{l\in\mathcal{L}}q_l = N_\mathrm{p},\\
        & q_l \geq 0.
    \end{align}
\end{subequations}

To simplify the analytical tractability, by removing the logarithmic function and focusing on the argument $\gamma$, the optimization problem \eqref{opt:outage_SE_minimization} can be reformulated as 
\begin{subequations}\label{opt:outage_SNR_minimization}
    \begin{align}
        \underset{q_l,~\forall l\in\mathcal{L}}{\mathrm{minimize}}
        \quad& \mathrm{Pr}\{\gamma \leq \gamma_\mathrm{th}\},\label{obj:outage_SNR_minimization}\\
        \mathrm{subject\:to}
        \quad
        & \sum_{l\in\mathcal{L}}q_l = N_\mathrm{p},\\
        & q_l \geq 0,
    \end{align}
\end{subequations}
where $\gamma_\mathrm{th}$ is the target \ac{RSNR}, defined as $\gamma_\mathrm{th}\triangleq  2^{\xi_\mathrm{th}} - 1$.
The outage probability in the objective function \eqref{obj:outage_SNR_minimization} is equivalent to the \ac{CDF} of the \ac{RSNR} in \eqref{eq:CDF_totalSNR} at $\gamma = \gamma_\mathrm{th}$ as 
\begin{align}\label{eq:outage_totalSNR}
    \mathrm{Pr}\{\gamma \leq \gamma_\mathrm{th}\} = \int_{-\infty}^{\gamma_\mathrm{th}} f_{\Gamma}(t)dt = F_\Gamma(\gamma_\mathrm{th}).
\end{align}

By utilizing the CDF of the \ac{RSNR} derived in \eqref{eq:CDF_totalSNR}, the outage probability for any target \ac{RSNR} can be obtained in closed form without using the Monte Carlo computation, unlike the conventional approaches~\cite{blockage_iimori,blockage_uchimura,blockagehybrid_uchimura,blockagewcnc_uchimura}.
For the sake of notation convenience, the outage probability in \eqref{eq:outage_totalSNR} is expressed as a function of the panel allocation vector $\mathbf{q}$, denoted as 
\begin{align}
    f_\mathrm{out}(\mathbf{q}) = F_{\mathrm{\Gamma}}(\gamma_\mathrm{th}|\mathbf{q}).
\end{align}

To solve the optimization problem in \eqref{opt:outage_SNR_minimization}, we propose a panel allocation algorithm based on brute-force search.
The pseudocode of the proposed algorithm is provided in \algref{alg:brute-force}.

%%% Outage minimization %%%
\begin{algorithm}[t]
    \caption{Outage Probability Minimization}\label{alg:brute-force}
    \textbf{Input:~}$\xi_\mathrm{th}$, $\gamma_\mathrm{tx}$, $N_\mathrm{a}$, $N_\mathrm{t}$, $\kappa$, $L$, $p_\mathrm{blk}$
    \\
    \textbf{Output:} $\mathbf{q}^\mathrm{opt}\in\mathbb{Z}^L$
    \begin{algorithmic}[1]
    \STATE {Generate $\tilde{\mathbf{Q}}\in\mathbb{Z}^{L\times C}$ that contains all pattern vectors}
    \STATE {Calculate $\mathbf{q}^\mathrm{opt}$ by solving the problem \eqref{opt:allocation_last}}
    \end{algorithmic}
\end{algorithm}

The proposed algorithm in \algref{alg:brute-force} determines the optimal panel allocation by exhaustively searching all possible combinations of panels and paths.
Although communication outage is unavoidable for any panel allocation when all utilized paths are simultaneously blocked, the proposed algorithm can reduce the probability of such complete blockage while maintaining the achievable \ac{SE} by appropriately aligning beams toward multiple paths.
The number of total patterns for panel allocation is given by
\begin{align} \label{eq:num_pattern}
    C=\sum_{q_1=1}^{N_\mathrm{p}}\binom{N_\mathrm{p}+L-q_1-2}{L-2}.
\end{align}
\par
Let us define the panel allocation matrix including all patterns as $\tilde{\mathbf{Q}} = [\tilde{\mathbf{q}}_1, \ldots, \tilde{\mathbf{q}}_C] \in \mathbb{C}^{L \times C}$, where $\tilde{\mathbf{q}}_i$ represents the $i$-th panel allocation candidate among total $C$ patterns.
Using the panel allocation matrix $\tilde{\mathbf{Q}}$, the optimal panel allocation vector $\mathbf{q}^\mathrm{opt}\in\mathbb{Z}^{L}$ can be calculated as
\begin{align}\label{opt:allocation_last}
    \mathbf{q}^\mathrm{opt} = \underset{\tilde{\mathbf{q}}_i \in \tilde{\mathbf{Q}}}{\mathrm{argmin}}~ f_\mathrm{out}(\tilde{\mathbf{q}}_i).
\end{align}

\subsection{Outage Minimization with Reinforcement for Average SE}
\label{subsec:outmin_ASE}
The proposed algorithm in \algref{alg:brute-force}, described in the previous subsection, considers only minimizing the outage probability.
Focusing only on minimizing the outage probability may lead to a significant degradation in the average \ac{SE}, as will be demonstrated in Section~\ref{sec:results}.
To address this challenge, we enhance the previous algorithm by introducing a modification to mitigate the degradation of the average \ac{SE}.
The pseudocode of this enhanced algorithm is provided in \algref{alg:outmin_ASE}.

Let $\tilde{p}_i$ and $p^\mathrm{min}$ denote the outage probability corresponding to the $i$-th panel allocation vector $\tilde{\mathbf{q}}_i$ and the minimum outage probability, which are given by
\begin{align}
    \label{eq:outprob_ith_candidate}
    \tilde{p}_i &= f_\mathrm{out}(\tilde{\mathbf{q}}_i),\quad \forall i \in \{1,2, \ldots, C\}, \\
    \label{eq:outprob_min}
    p^\mathrm{min} &= \underset{i \in \{1,2, \ldots, C\}}{\mathrm{min}} \ \tilde{p}_i.
\end{align}

These outage values in \eqref{eq:outprob_ith_candidate}, \eqref{eq:outprob_min} can be obtained by \algref{alg:brute-force}. 
To enhance the average SE performance, the proposed algorithm determines the optimal allocation vector $\mathbf{q}^\mathrm{opt}$ that maximizes the average \ac{RSNR} (which corresponds to maximizing the upper bound of the average SE in \eqref{eq:average_SE_Jensen}) from the allocation candidates such that the absolute difference from the minimum outage probability $p^\mathrm{min}$ is less than or equal to a small value $\varepsilon$. 
The optimization problem is formulated as
\begin{subequations}
\label{eq:out_min_ase}
\begin{align}
    & \mathbf{q}^\mathrm{opt} = \underset{j\in\mathcal{J}}{\mathrm{argmax}}~~f_\mathrm{ave}^\gamma( \tilde{\mathbf{q}}_j ), \\
    & \text{subject to} \  \mathcal{J} = \left \{j \Big | ~ |p^\mathrm{min} - \tilde{p}_{j}| \leq \varepsilon \right \},
\end{align}
\end{subequations}
where $f_\mathrm{ave}^\gamma(\mathbf{q})$ is the average RSNR from \eqref{eq:totalSNR_expectation}, which is expressed as 
\begin{align}\label{eq:RSNR_expectation_panel}
    f_\mathrm{ave}^\gamma(\mathbf{q}) = \dfrac{\gamma_\mathrm{tx}N_\mathrm{a}^2 \left(1 - p_\mathrm{blk}\right)}{N_\mathrm{t}(\kappa+1)(L-1)} \left(\kappa(L-1) q_1^2 + q_2^2 + \dots + q_L^2\right).
\end{align}

This optimization can also be solved through brute-force search, as in the optimization problem \eqref{opt:allocation_last} in the previous subsection.

\begin{algorithm}[t]
    \caption{Outage Probability Minimization with Reinforcement for Average SE}\label{alg:outmin_ASE}
    \textbf{Input:~}$\xi_\mathrm{th}$, $\gamma_\mathrm{tx}$, $N_\mathrm{a}$, $N_\mathrm{t}$, $\kappa$, $L$, ${p}_\mathrm{blk}$, $\varepsilon$
    \\
    \textbf{Output:} $\mathbf{q}^\mathrm{opt}\in\mathbb{Z}^L$
    \begin{algorithmic}[1]
    \STATE Generate $\tilde{\mathbf{Q}}\in\mathbb{Z}^{L\times C}$ that contains all pattern vectors
    \STATE Calculate $\tilde{p}_i$ and $p^\mathrm{min}$ from \eqref{eq:outprob_ith_candidate} and \eqref{eq:outprob_min}
    \STATE Calculate $\mathbf{q}^\mathrm{opt}$ by solving the problem \eqref{eq:out_min_ase}
    \end{algorithmic}
\end{algorithm}

\subsection{Evaluation of Computational Complexity}
\label{subsec:complexty}
As described in the previous subsection, the proposed algorithm exhaustively searches all patterns of panel allocation based on the brute-force search.
From \eqref{eq:num_pattern} and \algref{alg:brute-force} , the worst-case complexity of the panel allocation algorithm is $\mathcal{O} (\binom{N_\mathrm{p}+L-2}{L-1} 2^{\mathrm{min}(L, N_\mathrm{p})})$.
Thus, combinatorial explosion may occur when the numbers of panels $N_\mathrm{p}$ and paths $L$ become large.
However, in practical \ac{mmWave} scenarios, these parameters typically remain relatively small, \textit{e.g.}, on the order of 1 to 10~\cite{blockage_modeling_akdeniz,mmwave_numpaths_hanzo}.
Therefore, the optimization problem remains computationally feasible even with brute-force search.
In this paper, brute-force search is adopted to identify the theoretical limit of the proposed panel allocation through the globally optimal solution, although the formulation can be extended to low-complexity solvers such as meta-heuristic methods.
Thus, brute-force search is adopted for panel allocation.

\par
Since the dominant factor in the computational complexity is the total number of panel allocation patterns $C$, we evaluate $C$ under various parameter settings for $N_\mathrm{p}$ and $L$.
\figref{fig:num_pattern} shows the results, where the number of antenna elements per panel is set to $N_\mathrm{a}=32$ and the total number of antenna elements is $N_\mathrm{t}=N_\mathrm{a}N_\mathrm{p}\in[64,512]$.

As shown in the figure, the number of allocation patterns is at most $10^6$ under practical parameter settings. 
Thus, the optimization problem can be solved with reasonable complexity in the considered mmWave system.
Furthermore, the proposed panel allocation algorithms require only long-term statistics, such as the number of paths and the Rician $K$-factor, without the need for instantaneous CSI.
Therefore, it is not required to solve the optimization problem at every channel coherence time, which significantly mitigates the complexity requirements in terms of implementation.
Moreover, since the panel allocation is updated only according to long-term statistical variations rather than at every channel coherence time, the computational burden can be substantially relaxed in practical implementation.

\begin{figure}[t]
    \begin{center}
        \includegraphics[width=\linewidth]{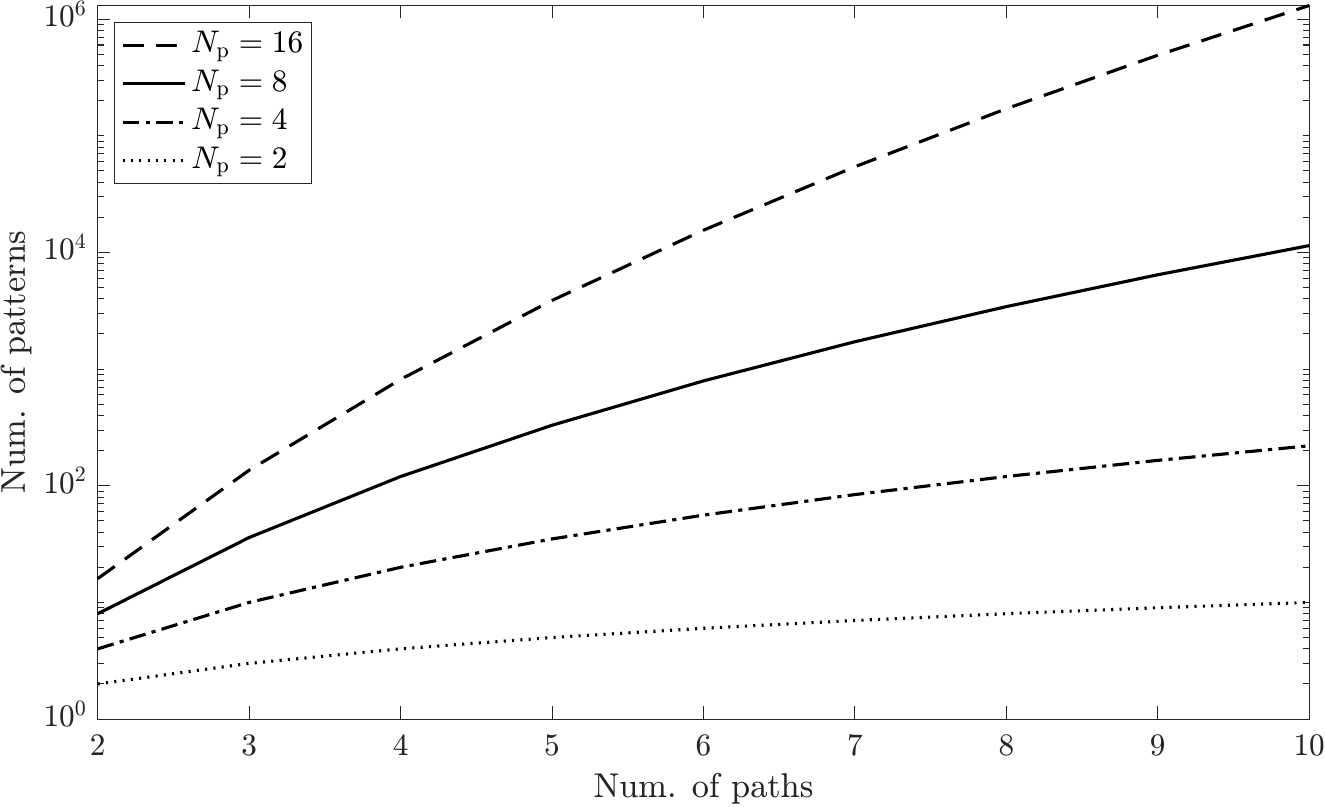}
    \end{center}
    \vspace*{-0.2cm}
    \caption{The total number of allocation patterns $C$ as a function of $L$. The number of antennas per panel is $N_\mathrm{a}=32$, and the total number of antennas is $N_\mathrm{t}=N_\mathrm{a} N_\mathrm{p} \in [64, 512]$.}
    \label{fig:num_pattern}
\end{figure}

\section{Performance Assessment}\label{sec:results}
This section evaluates the performance of the proposed analog beamforming designs. 
The simulation parameters used in this paper are as follows.
The carrier frequency $f_\mathrm{c}$ is set to $28~\mathrm{GHz}$, and the system bandwidth $W$ is $100~\mathrm{MHz}$.
The Rician $K$-factor $\kappa$ is assumed to be $10~\mathrm{dB} $\cite{kfactor_samimi}, and the number of paths $L$ is set to $4$.
The path blockage probability $\hat{p}_l$ is set to $\hat{p}_l \sim \mathcal{U}(0.2,0.6)$~\cite{pathloss_rappaport,pblk_raghavan} with the average blockage probability $p_\mathrm{blk} = 0.4$.
This simulation considers a multi-panel array configuration, where the number of panels $N_\mathrm{p}$ is $8$, and each panel consists of $N_\mathrm{a} = 32$ antenna elements.
The hyperparameter of \algref{alg:outmin_ASE} is set to $\varepsilon=0.05$.
Unless otherwise specified, the transmit SNR is set to $\gamma_\mathrm{tx}=10\ [\mathrm{dB}]$ in the following simulations.

\par
The system performance is evaluated by \ac{CDF} of \ac{SE}, average \ac{SE}, and outage probability of \ac{SE}.
Note that, the \ac{SE}, average \ac{SE}, and outage probability are defined in \eqref{eq:SE_difinition}, \eqref{eq:average_SE_Jensen} and \eqref{eq:outage_probability}, respectively.

To clarify the effectiveness of the proposed beamforming designs, the following four methods are assessed in the simulation.
\begin{itemize}
    \item[1)] \textbf{\textit{LoS Concentration}} : 
        This method is the commonly used beamforming design that generates a sharp beam aligned with the \ac{LoS} path between the \ac{BS} and \ac{UE}. 
        This beamforming design is adopted in current systems, where all panels are coordinated to direct the beam toward a single LoS path~\cite{3GPP_FDMIMO}.
    \item[2)] \textbf{\textit{Uniform Path Allocation}} : 
        This method allocates panels equally across all paths, not only including \ac{LoS} path but also \ac{NLoS} paths.
        This naive panel allocation is used in the conventional works~\cite{mmWaveReliable_Kumar,BeamSwitch_Wilson,mp_terui}.
    \item[3)] \textbf{\textit{OutMin}} : 
        This method is the proposed approach presented in \algref{alg:brute-force}, which focuses solely on minimizing the outage probability.
    \item[4)] \textbf{\textit{OutMin w/ Average SE}} : 
        This method is the proposed approach presented in \algref{alg:outmin_ASE}, which considers both the outage probability and the average SE.
\end{itemize}

Hereafter, numerical evaluations are provided based on the aforementioned metrics.
Section~\ref{subsec:CDF_SE} validates the analytical \ac{CDF} against Monte Carlo simulations.
Section~\ref{subsec:sim_panel allocation} presents the optimized panel allocations, and Section~\ref{subsec:sim_outage} evaluates the outage-probability and average \ac{SE} trade-off as a function of the target \ac{SE}.
Sections~\ref{subsec:sensitivity_numpaths} and~\ref{subsec:sensitivity_epsilon} provide sensitivity analyses with respect to the number of paths and the hyperparameter $\varepsilon$, respectively.
Finally, Section~\ref{subsec:sim_average_SE} evaluates the average \ac{RSNR} and \ac{SE} over a range of transmit \ac{SNR}.

\subsection{Theoretical Analyses for the CDF of the SE}
\label{subsec:CDF_SE}
% Figure 
\begin{figure*}[t]
    \begin{minipage}[b]{0.5\textwidth}
        \centering
        \includegraphics[width=\linewidth]{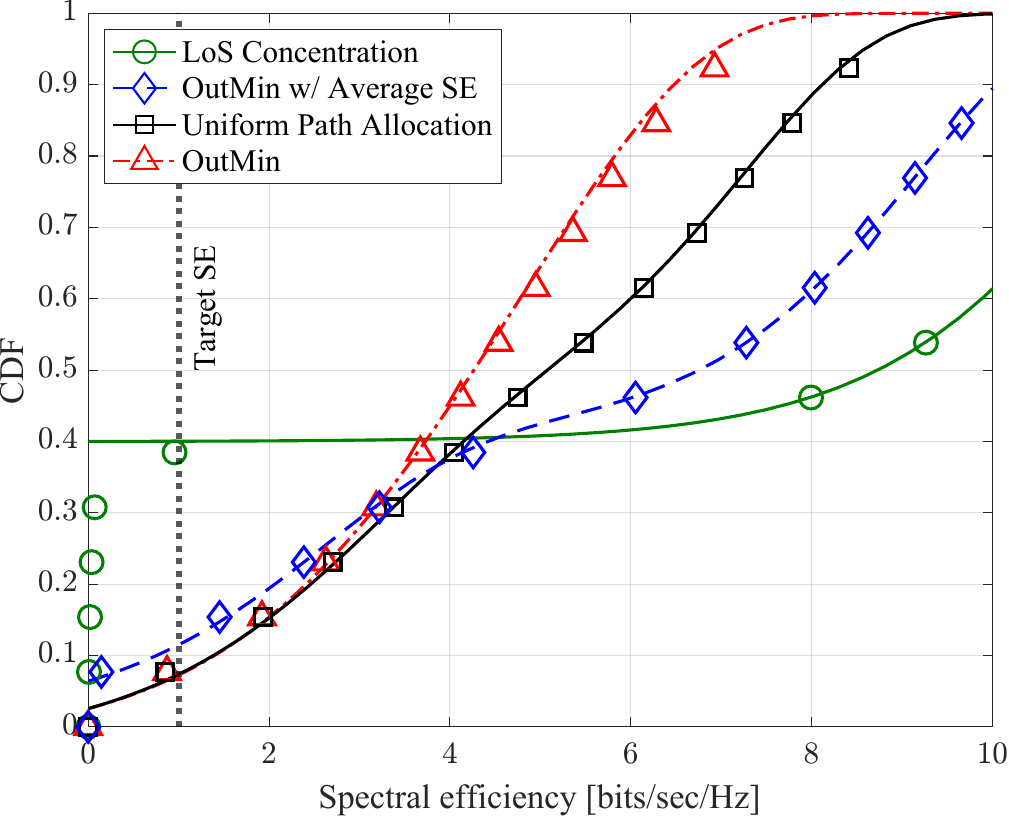}
        \vspace*{-0.2cm}
        \subcaption{Target SE: $\xi_\mathrm{th}$ = 1 [bits/sec/Hz].}
        \label{fig:CDF_SE_TargetSE=1}
    \end{minipage}
    % \hspace{5mm}
    % 
    \begin{minipage}[b]{0.5\textwidth}
        \centering
        \includegraphics[width=\linewidth]{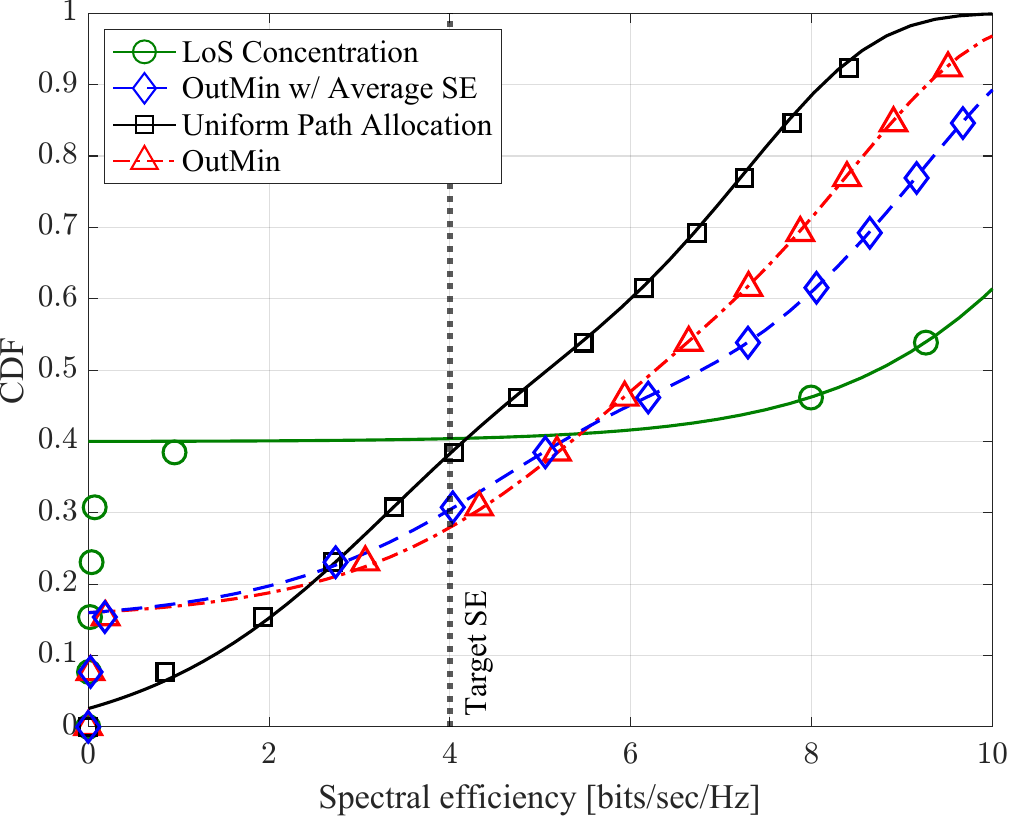}
        \vspace*{-0.2cm}
        \subcaption{Target SE: $\xi_\mathrm{th}$ = 4 [bits/sec/Hz].}
        \label{fig:CDF_SE_TargetSE=4}
    \end{minipage}
    \caption{CDF of the SE for Multiple Target SE. The solid lines represent the analytically calculated \acp{CDF}, while the markers indicate the simulated \acp{CDF}.}
    \label{fig:CDF_SE}
\end{figure*}
\begin{figure*}[t]
    %% Brute Force
    \begin{minipage}[b]{0.5\textwidth}
        \begin{center}
            \includegraphics[width=\textwidth]{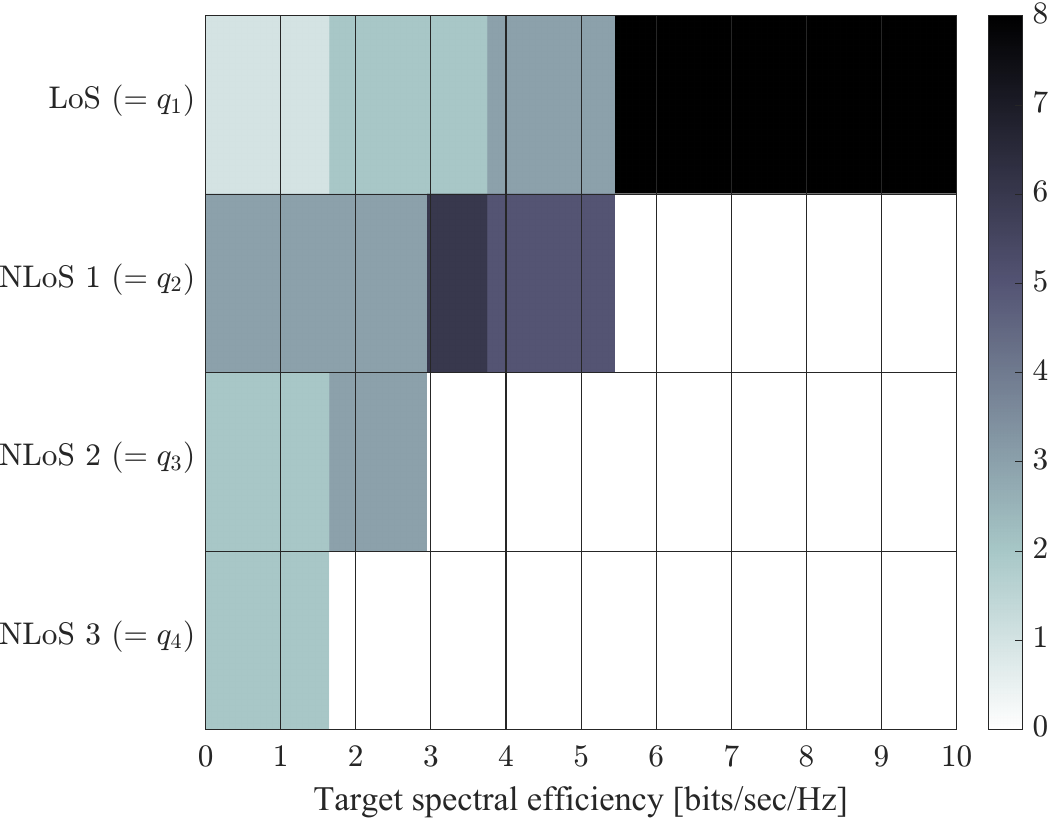}
        \end{center}
        \vspace*{-0.2cm}
        \subcaption{\algref{alg:brute-force} \textit{OutMin}}
        \label{fig:heatmap_panels_bruteforce}
    \end{minipage}
    \hspace{1mm}
    % 
    %% Brute Force + Threshold
    \begin{minipage}[b]{0.5\textwidth}
        \begin{center}
            \includegraphics[width=\textwidth]{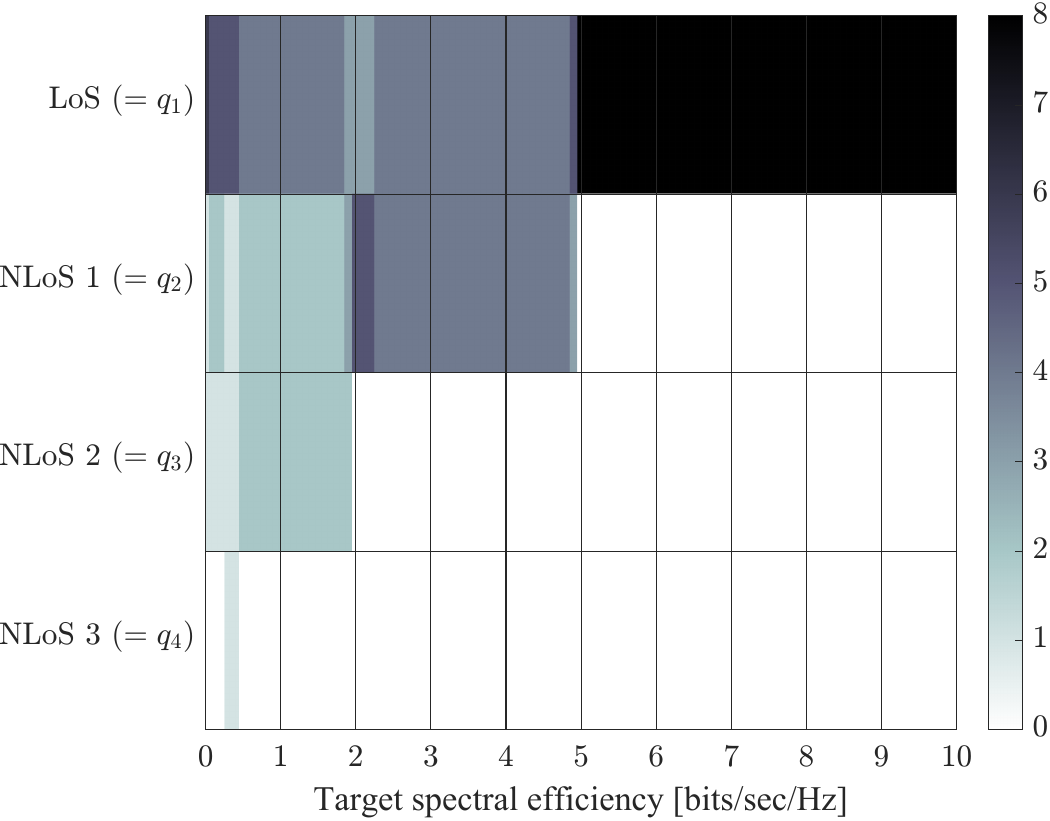}
        \end{center}
        \vspace*{-0.2cm}
        \subcaption{\algref{alg:outmin_ASE} \textit{OutMin w/ Average SE} ($\varepsilon=0.05$)}
        \label{fig:heatmap_panels_outminASE}
    \end{minipage}
    \caption{
    % \Kabuto{
    Panel allocation results in the proposed methods.
    The vertical axis represents the number of panels aligned to each path, corresponding to each element of $\mathbf{q}^\mathrm{opt}$ (\textit{e.g.}, $q_1=3$ and $q_2=1$ indicate that the number of panels aligned to the LoS and NLoS 1 are three and one, respectively)
    % }
    }
    \label{fig:heatmap_panels}
\end{figure*}

We evaluate the behavior of the \ac{CDF} of the \ac{SE} by comparing the analytically calculated \ac{CDF} with the numerically computed \ac{CDF} from Monte Carlo simulations with realistic model parameters.
The \acp{CDF} are shown in \figref{fig:CDF_SE_TargetSE=1} and \figref{fig:CDF_SE_TargetSE=4}, where the target \ac{SE} values are set to $\xi_\mathrm{th} = 1$ and $4~[\mathrm{bits/sec/Hz}]$, respectively.
The solid lines represent the analytically calculated \acp{CDF}, while the markers indicate the numerically computed \acp{CDF}.
As shown in the figures, the theoretical results closely match the numerical results.
This confirms the validity of the approximations used in the theoretical analysis.
However, in regions with low \ac{SE}, slight discrepancies arise between the theoretical and numerical results due to the approximations employed in the theoretical model in \eqref{eq:blockage_PMF} and \eqref{eq:array_response_approx}.
In the theoretical analysis, we assume that the beamwidth of the main lobe approaches zero.
This simplification implies that only a single path can fall within the main lobe.
In practice, however, multiple paths may exist within the same beam, which unintentionally increases the received signal power.
Furthermore, the theoretical analysis neglects the effect of sidelobes, whereas in the practical simulation, the energy collected through sidelobes can also be exploited, leading to further performance improvements.
Consequently, as illustrated in \figref{fig:CDF_SE}, the \ac{SE} performance of the simulation is slightly superior to the theoretical results.
From these observations, it can be confirmed that the two approximations used in the theoretical analysis provide conservative predictions that assume worst-case performance.
Accordingly, the proposed method can be interpreted as providing a conservative and robust beamforming design, aiming to ensure stable communication under blockage environments.
Moreover, the impact of these approximations is mainly confined to the extremely low-\ac{SE} region and rapidly diminishes as the \ac{SE} increases.
Thus, they do not materially affect the evaluation of the proposed robust beamforming design.

Compared to the proposed methods, the conventional method, \textit{LoS Concentration}, has a high probability (approximately 0.4) of the SE being $0~[\mathrm{bits/sec/Hz}]$, which is equivalent to the average blockage probability $p_\mathrm{blk}$.
This result indicates that focusing only on the LoS path results in frequent disconnections whenever the LoS path is blocked, leading to unstable communication.
In contrast, the proposed methods and \textit{Uniform Path Allocation}, which utilize multiple paths, can reduce the probability of the SE being $0~[\mathrm{bits/sec/Hz}]$.
Especially in \textit{Uniform Path Allocation}, the probability becomes $p_\mathrm{blk}^L = 0.0256$ since this method utilizes all $L$ paths.
Therefore, unless all the paths are blocked, the communication is not disconnected.
This is also verified by the theoretical CDF expressed in \eqref{eq:CDF_totalSNR}, where the probability of the SE $\xi$ being zero (\textit{i.e.}, the RSNR $\gamma$ being zero) is given as $p_\mathrm{blk}^L$ when $N_\mathrm{b}=L$.

Furthermore, the proposed method, \textit{OutMin}, achieves the lowest \ac{CDF} value at the target \ac{SE}, demonstrating its flexibility in beamforming design to handle path blockages.
Focusing on \figref{fig:CDF_SE_TargetSE=1}, it is observed that the performance of \textit{OutMin} deteriorates at high \ac{SE}.
This degradation arises because the beams designed by \textit{OutMin} focus solely on minimizing the outage probability without considering the average SE.
In comparison, \textit{OutMin w/ Average SE} can suppress the degradation of the average \ac{SE} by incorporating mechanisms to enhance the average \ac{SE} as described in Section~\ref{subsec:outmin_ASE}.

% Figure 
\begin{figure*}[t]
    \begin{minipage}[b]{0.5\textwidth}
        \begin{center}
            \includegraphics[width=\linewidth]{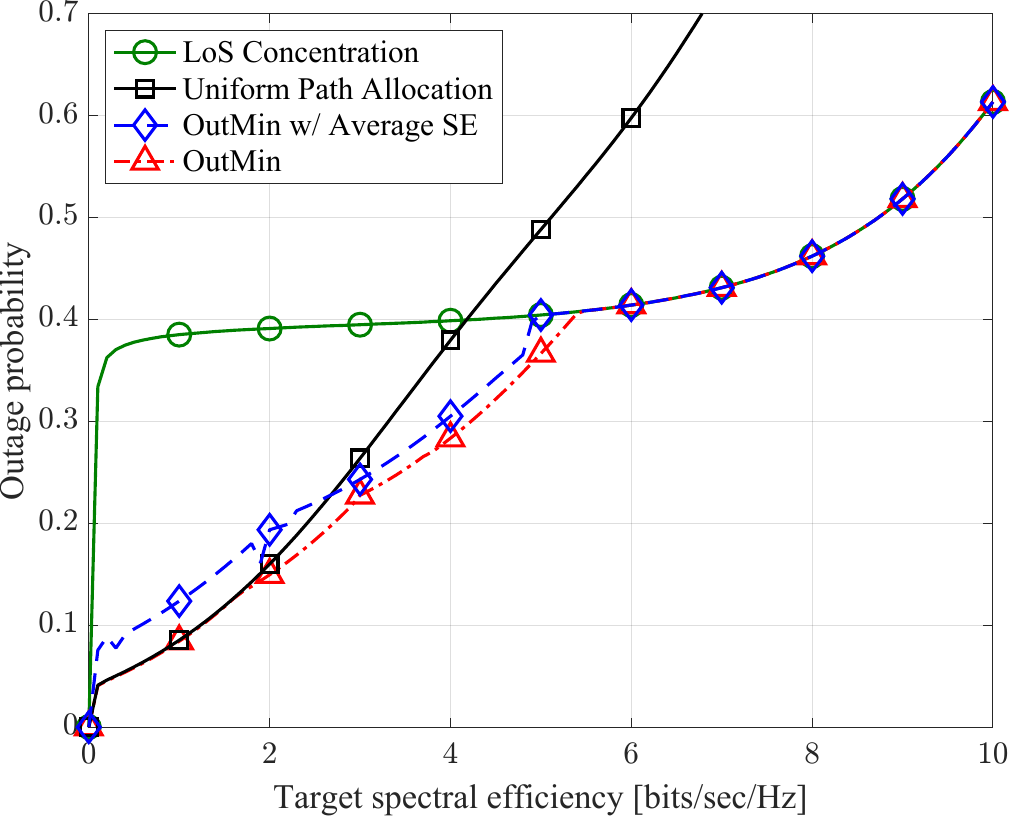}
        \end{center}
        \vspace*{-0.2cm}
        \caption{Outage probability as a function of target SE $\xi_\mathrm{th}$}
        \label{fig:TargetSE_OutProb}
    \end{minipage}
    \hspace{1mm}
    \begin{minipage}[b]{0.5\textwidth}
        \begin{center}
            \includegraphics[width=\linewidth]{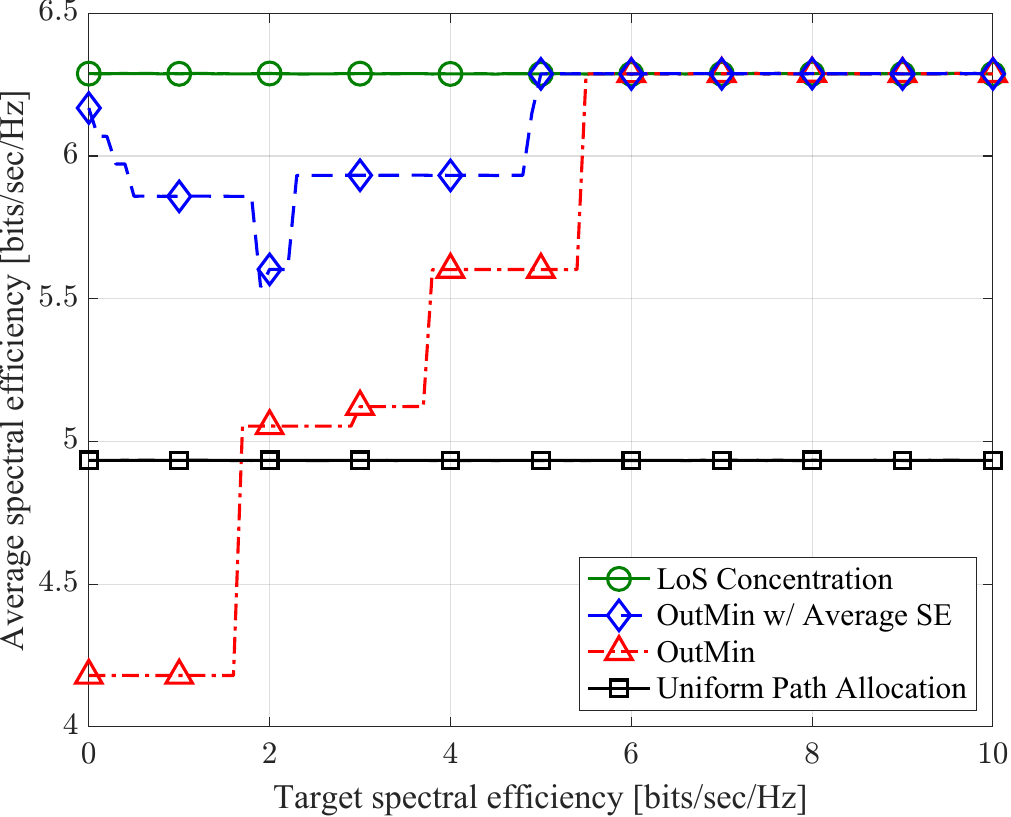}
        \end{center}
        \vspace*{-0.2cm}
        \caption{Average SE as a function of target SE $\xi_\mathrm{th}$}
        \label{fig:TargetSE_AveSE}
    \end{minipage}
\end{figure*}
\subsection{Panel Allocation Results}
\label{subsec:sim_panel allocation}
To visualize the optimized panel allocation under various target SE values, 
we present, in \figref{fig:heatmap_panels}, the panel allocation results in the proposed methods, (a) \textit{OutMin} and (b) \textit{OutMin w/ Average SE}.
In the figure, the vertical axis represents the number of panels aligned to each path, corresponding to each element of $\mathbf{q}^\mathrm{opt}$ (\textit{e.g.}, $q_1=3$ and $q_2=1$ indicate that the number of panels aligned to the LoS and NLoS 1 are three and one, respectively).
As shown in the figure, the proposed algorithms generate beams depending on the target SE. In the high target SE region, a single sharp beam focused solely on the LoS path is designed, whereas in the low target SE region, multiple beams aligned with both LoS and NLoS paths are designed.
Comparing \textit{OutMin} in \figref{fig:heatmap_panels_bruteforce} and \textit{OutMin w/ Average SE} in \figref{fig:heatmap_panels_outminASE}, both methods utilize multiple paths, however, differences are observed in the number of panels allocated to each path.
In \textit{OutMin}, more panels are allocated to \ac{NLoS} paths than to the \ac{LoS} path.
This is because \textit{OutMin} suppresses the imbalance in the equivalent path gains $g_{q_l}$ for each path in \eqref{eq:heq_approx} to obtain the higher spatial diversity.
In contrast, \textit{OutMin w/ Average SE} allocates more panels to the \ac{LoS} path than to \ac{NLoS} paths to improve the average \ac{SE} by tolerating some degradation in the outage probability.
For the same target \ac{SE}, \textit{OutMin w/ Average SE} tends to reduce the number of beams $N_\mathrm{b}$ compared to \textit{OutMin} to obtain high-gain beams for fewer paths rather than to achieve spatial diversity.

%%%%%%%%%%%%%%%%%%%%%%%%%%%%%%%%%%%%%%%%%%%%%%
\subsection{Outage Probability and Average SE vs. Target SE}
\label{subsec:sim_outage}
To evaluate the outage probability defined in \eqref{eq:outage_probability}, \figref{fig:TargetSE_OutProb} shows the outage probability as a function of the target \ac{SE} $\xi_\mathrm{th}$.
For low target \ac{SE} values, the conventional method, \textit{LoS Concentration}, experiences outages with a probability equal to the path blockage probability $p_\mathrm{blk}=0.4$ due to \ac{LoS} blockage.
In contrast, the proposed methods can suppress the outage probability by exploiting multiple paths.
Notably, in the region where the target \ac{SE} is approximately $\xi_\mathrm{th} \geq 3~[\mathrm{bits/sec/Hz}]$, the proposed algorithms in \algref{alg:brute-force} reduce the outage probability effectively compared to \textit{Uniform Path Allocation} owing to utilizing the LoS path as shown in Fig.~\ref{fig:heatmap_panels}. 
The proposed method using \algref{alg:outmin_ASE} allows an outage probability degradation of up to $\varepsilon$ to improve the average \ac{SE}.
As a result, the outage probability is degraded by at most $\varepsilon$ compared to the result of \algref{alg:brute-force}.
To achieve a high target \ac{SE}, beams must be concentrated on the \ac{LoS} path. 
Thus, \textit{LoS Concentration} achieves the lowest outage probability at the high target \ac{SE} region.
The proposed methods achieve spatial diversity in the low target \ac{SE} region by utilizing multiple beams aligned with multiple paths, and obtain array gain in the high target \ac{SE} region by focusing beams on the LoS path, thereby suppressing the outage probability for all target SE values.

To evaluate the average \ac{SE} performance under various target \ac{SE} values $\xi_\mathrm{th}$, \figref{fig:TargetSE_AveSE} illustrates the variation of the average \ac{SE} with respect to the target \ac{SE} $\xi_\mathrm{th}$.
The average \ac{SE} is evaluated as $ \mathbb{E}[\xi] = \mathbb{E}[ \log_2(1 + \gamma)]$ defined in \eqref{eq:average_SE_Jensen}.
As shown in the figure, \textit{LoS Concentration} achieves the highest average \ac{SE} for any target \ac{SE}, while \textit{Uniform Path Allocation}, which utilizes multiple paths, fails to achieve improvements in the averaged \ac{SE}.
In contrast, since the proposed methods optimize panel allocation to adapt to the target \ac{SE}, the average \ac{SE} varies with the target \ac{SE} and improves at higher target \ac{SE} values.
Comparing the outage probability in \figref{fig:TargetSE_OutProb} and the average \ac{SE}  in \figref{fig:TargetSE_AveSE}, it can be seen that there is a trade-off between them in the low target \ac{SE}  region, where the outage probability can be improved at the cost of the average \ac{SE}  by designing multiple beams.
On the other hand, in the high target \ac{SE}  region, the average \ac{SE} improves by concentrating beams only on the LoS path to obtain array gain, resulting in the improvement of the outage probability.

\begin{figure}[t]
    \begin{center}
        \includegraphics[width=\linewidth]{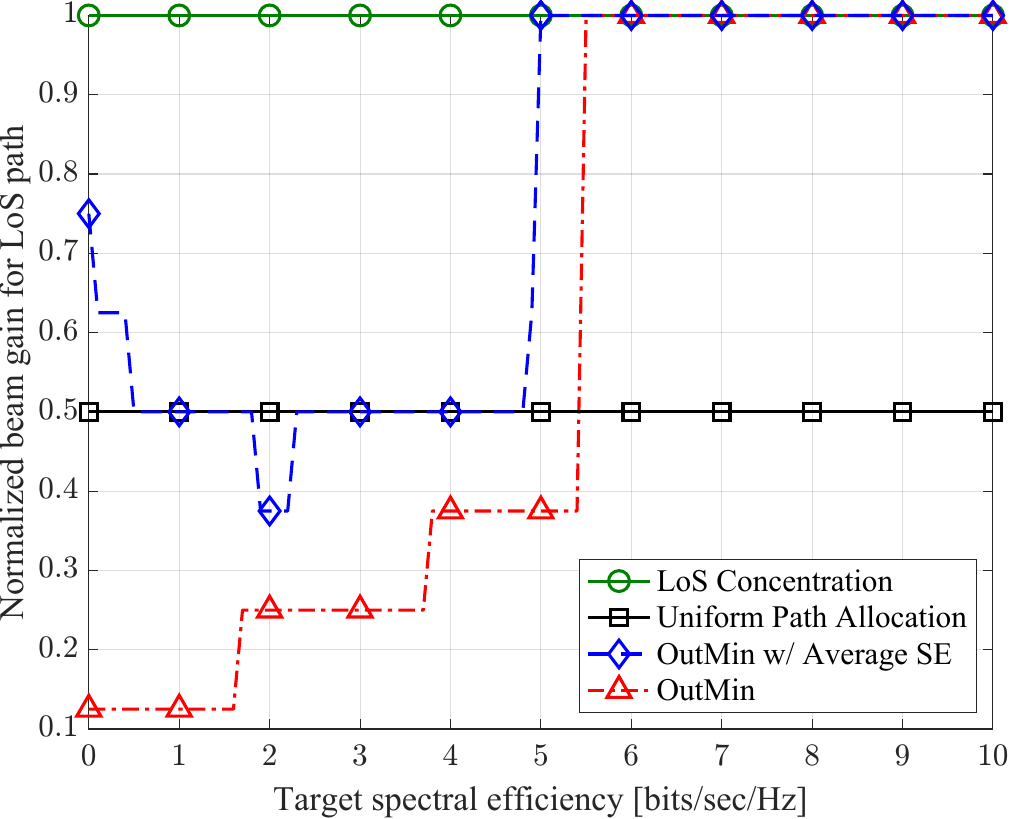}
    \end{center}
    \vspace*{-0.2cm}
    \caption{Normalized beam gain $G_\mathrm{LoS}$ for LoS path as a function of $\xi_\mathrm{th}$}
    \label{fig:TargetSE_GainRatio}
\end{figure}

\par
In \figref{fig:TargetSE_AveSE}, \textit{OutMin w/ Average SE} exhibits a non-monotonic variation in the average \ac{SE}.
To clarify the cause of this behavior, we present \figref{fig:TargetSE_GainRatio}, which illustrates the normalized beam gain for the \ac{LoS} path $G_\mathrm{LoS}$ as a function of the target \ac{SE} $\xi_\mathrm{th}$. 
$G_\mathrm{LoS}$ is defined as 
\begin{align} \label{eq:r_LoS}
    G_\mathrm{LoS} \triangleq \frac{q_1 N_\mathrm{a}}{\sum_{i=1}^{L} q_i N_\mathrm{a}} = \frac{q_1}{N_\mathrm{p}},
\end{align}
where $q_i$ is the $i$-th element of the panel allocation vector $\mathbf{q}$, and $q_1$ means the number of panels allocated to the \ac{LoS} path.
Since the beam gain for the $i$-th path is expressed as $N_\mathrm{a} q_i$, the normalized beam gain for the LoS path can be expressed as \eqref{eq:r_LoS}.
In the region with low target SE ($\xi_\mathrm{th} \leq 2$), it is necessary to utilize multiple paths including both LoS and NLoS paths to achieve the target SE.  
If the target SE is extremely low, this target SE can be achieved with high probability by utilizing multiple paths, even when most of beams are concentrated on the LoS path.
However, as the target SE increases, the required beam gain for NLoS paths to achieve the target SE increases.
As a result, the normalized beam gain for the LoS path $G_\mathrm{LoS}$ decreases as the target SE $\xi_\mathrm{th}$ increases when $\xi_\mathrm{th} \leq 2$, as shown in Fig.~\ref{fig:TargetSE_GainRatio}.
In contrast, when the target SE becomes sufficiently high, the use of NLoS paths with small path gains are insufficient for achieving the high target SE, resulting in all beams being concentrated primarily on the LoS path.
Thus, when $\xi_\mathrm{th}\geq 2$, the normalized beam gain $G_\mathrm{LoS}$ increases with the target SE, converging to the result of \textit{LoS concentration}.

\subsection{Sensitivity Analysis for The Number of Paths $L$}
\label{subsec:sensitivity_numpaths}
To evaluate the performance with respect to the number of paths $L$, \figref{fig:sensitivity_numpaths} presents the outage probability and the average \ac{SE} as a function of $L$.
As shown in the figures, all methods, except for \textit{LoS Concentration}, exhibit degradation in both outage probability and average \ac{SE} as the number of paths increases.
In particular, \textit{Uniform Path Allocation}, which allocate array gain evenly to all paths, experiences significant performance degradation.
This is because, according to \eqref{eq:path_gain}, the gain of each \ac{NLoS} path relatively decreases as the number of paths increases, and consequently, uniformly distributing the array gain toward all paths causes the degradation of effective channel gain.
Nevertheless, the performance trends remain consistent regardless of the number of paths $L$, and \textit{OutMin} consistently achieves the minimum outage probability.
This is because \textit{OutMin} adaptively selects the optimal panel allocation from all feasible allocation candidates.
Furthermore, \textit{OutMin w/ Average SE} can improve the average \ac{SE} compared with \textit{OutMin} by adjusting the panel allocation through the hyperparameter $\varepsilon$ so as to more effectively exploit the \ac{LoS} path gain.
The selection of the hyperparameter $\varepsilon$ is discussed in the following subsection.

\begin{figure*}[t!]
    \begin{minipage}[b]{0.5\textwidth}
        \centering
        \includegraphics[width=\linewidth]{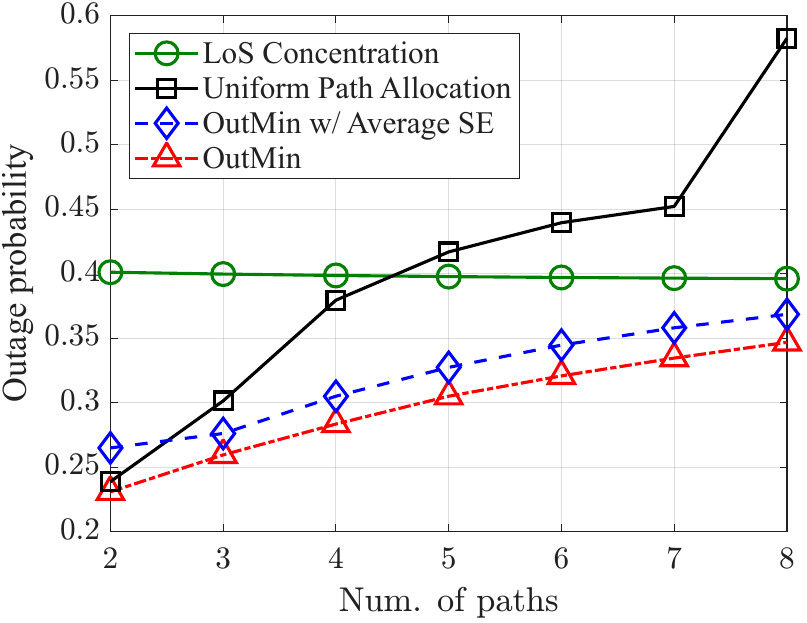}
        \vspace*{-0.2cm}
        \subcaption{Outage probability as a function of the number of paths $L$.}
        \label{fig:OutProb_numpaths}
    \end{minipage}
    % \hspace{5mm}
    % 
    \begin{minipage}[b]{0.5\textwidth}
        \centering
        \includegraphics[width=\linewidth]{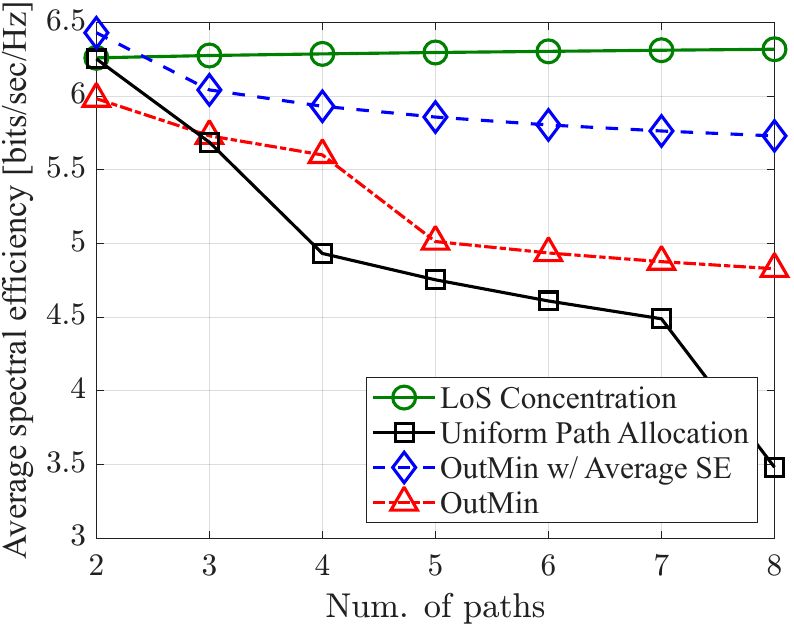}
        \vspace*{-0.2cm}
        \subcaption{Average \ac{SE} as a function of the number of paths $L$.}
        \label{fig:AverageSE_numpaths}
    \end{minipage}
    \caption{Sensitivity analysis for the number of paths $L$, where target SE $\xi_\mathrm{th} = 4~[\mathrm{bits/sec/Hz}]$.}
    \label{fig:sensitivity_numpaths}
\end{figure*}

\subsection{Sensitivity Analysis for Hyperparameter $\varepsilon$}\label{subsec:sensitivity_epsilon}
To evaluate the impact of the hyperparameter $\varepsilon$ used in ~\algref{alg:outmin_ASE}, \figref{fig:sensitivity_hyperparam} shows the outage probability and average \ac{SE} as functions of $\varepsilon$.
It can be seen that the performance of \textit{OutMin w/ Average SE} lies between that of \textit{OutMin} and \textit{LoS Concentration}.
When $\varepsilon$ is small, its performance is close to \textit{OutMin}, whereas for large $\varepsilon$ it approaches \textit{LoS Concentration}.

The sensitivity analysis shows that increasing $\varepsilon$ can improve the average \ac{SE} at the expense of a higher outage probability, indicating a trade-off between these two performance metrics.
The hyperparameter $\varepsilon$ represents the allowable increase in the outage probability relative to the minimum outage probability achieved by Alg.~1.
Therefore, $\varepsilon$ can be selected according to the outage probability requirement of the system.
If a maximum allowable outage probability is specified by the system, selecting the largest $\varepsilon$ within the allowable range is a reasonable choice for improving the average \ac{SE}.

\begin{figure*}[t!]
    \begin{minipage}[b]{0.5\textwidth}
        \centering
        \includegraphics[width=\linewidth]{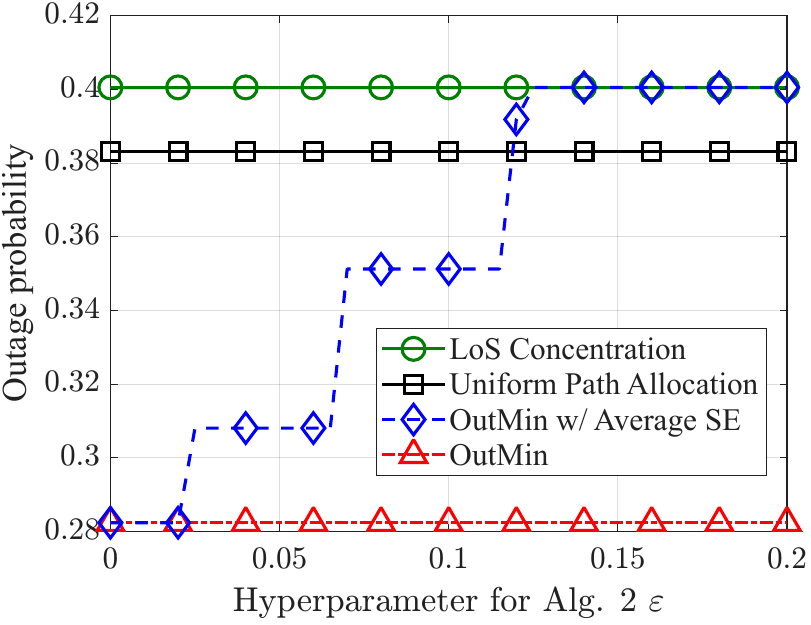}
        \vspace*{-0.2cm}
        \subcaption{Outage probability as a function of the hyperparameter $\varepsilon$.}
        \label{fig:OutProb_hyperparam}
    \end{minipage}
    % \hspace{5mm}
    % 
    \begin{minipage}[b]{0.5\textwidth}
        \centering
        \includegraphics[width=\linewidth]{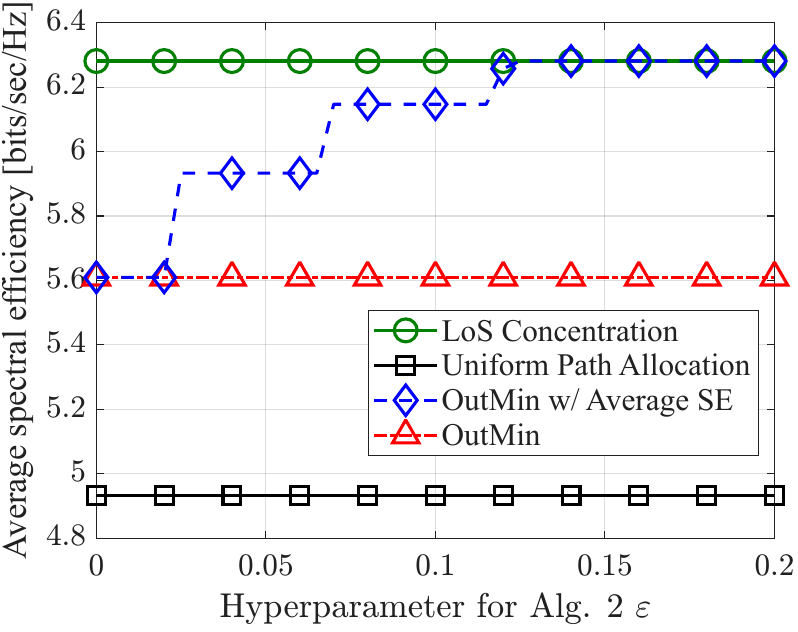}
        \vspace*{-0.2cm}
        \subcaption{Average \ac{SE} as a function of the hyperparameter $\varepsilon$.}
        \label{fig:AverageSE_hyperparam}
    \end{minipage}
    \caption{Sensitivity analysis for the hyperparameter $\varepsilon$, where target SE $\xi_\mathrm{th} = 4~[\mathrm{bits/sec/Hz}]$.}
    \label{fig:sensitivity_hyperparam}
\end{figure*}

%%%%%%%%%%%%%%%%%%%%%%

\subsection{Transmit SNR vs. Average RSNR and SE}
\label{subsec:sim_average_SE}
\begin{figure*}[t]
    %% Average RSNR
    \begin{minipage}[b]{0.5\textwidth}
        \begin{center}
        \includegraphics[width=\linewidth]{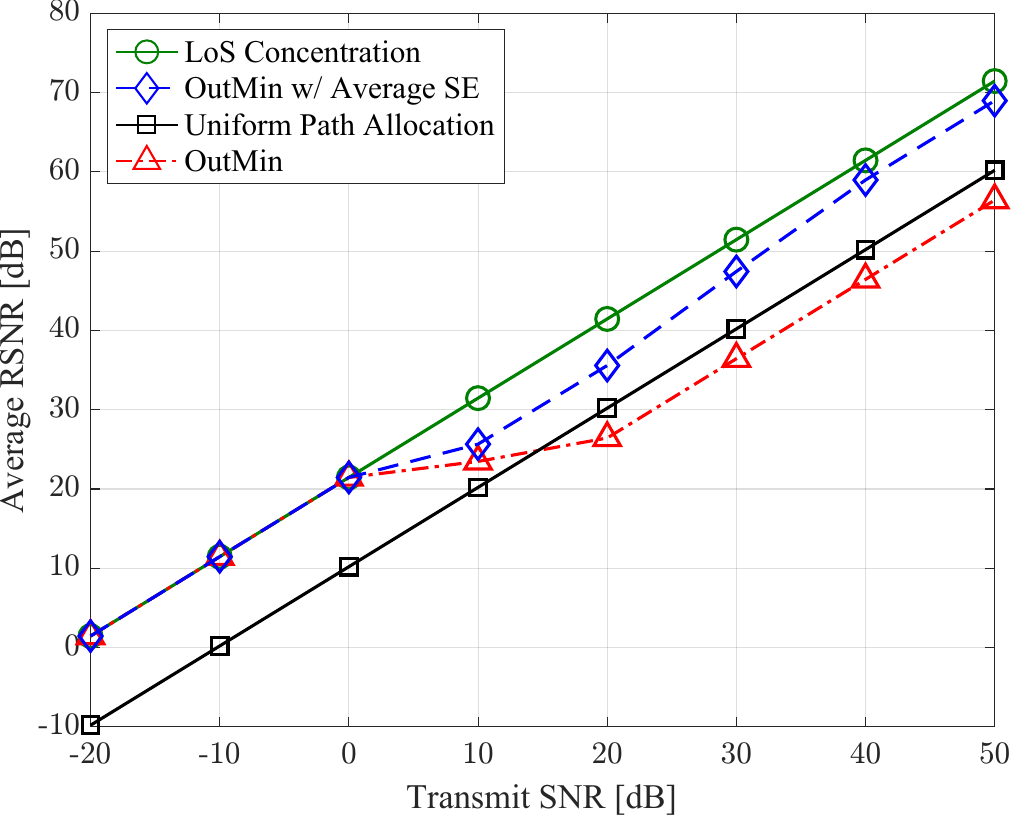}
        \end{center}
        \vspace*{-0.2cm}
        \caption{Average RSNR $\mathbb{E}[\gamma]$ as a function of transmit SNR $\gamma_\mathrm{tx}$: Target SE $\xi_\mathrm{th}$ = 4 [bits/sec/Hz]}
        \label{fig:TxSNR_AverageRSNR_TargetSE=4}
    \end{minipage}
    \hspace{1mm}
    %% Average SE
    \begin{minipage}[b]{0.5\textwidth}
        \begin{center}
        \includegraphics[width=\linewidth]{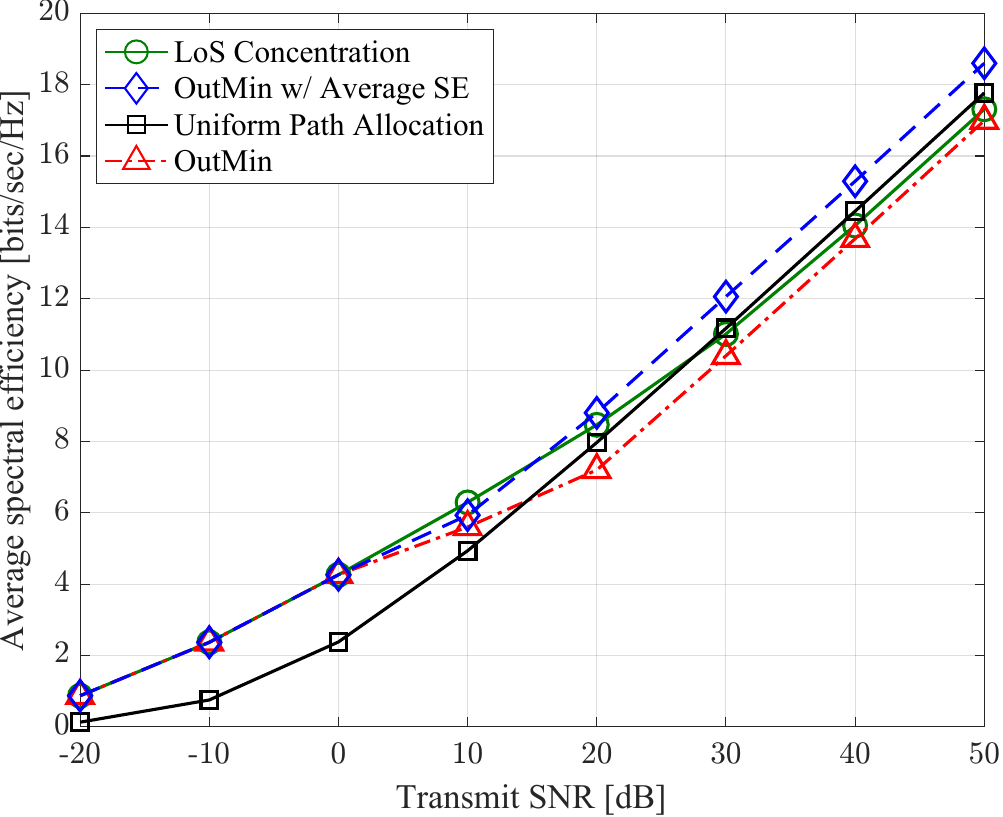}
        \end{center}
        \vspace*{-0.2cm}
        % \subcaption{Average SE $\mathbb{E}[\log_\mathrm{2}(1 + \gamma)]$}
        \caption{Average SE $\mathbb{E}[\xi]$ as a function of transmit SNR $\gamma_\mathrm{tx}$: Target SE $\xi_\mathrm{th}$ = 4 [bits/sec/Hz]}
        \label{fig:TxSNR_AverageSE_TargetSE=4}
    \end{minipage}
    % \vspace*{-0.5cm}
    % \caption{Average SE and RSNR as a function of transmit SNR $\gamma_\mathrm{tx}$: Target SE $\xi_\mathrm{th}$ = 4 [bits/sec/Hz].}
    \label{fig:TxSNR_AverageSE-RSNR_TargetSE=4}
\end{figure*}

To evaluate the average RSNR and the average SE under various transmit SNR $\gamma_\mathrm{tx}$, \figref{fig:TxSNR_AverageRSNR_TargetSE=4} and \figref{fig:TxSNR_AverageSE_TargetSE=4} present the average RSNR and \ac{SE} as a function of $\gamma_\mathrm{tx}$.
From \figref{fig:TxSNR_AverageRSNR_TargetSE=4}, it is observed that the conventional method, \textit{LoS Concentration}, achieves the highest average RSNR for all \ac{SNR} values.
This is because \textit{LoS Concentration} is the optimal solution of the average RSNR maximization \eqref{opt:average_SNR_dominant} as described in Section~\ref{subsec:average_SE}.
On the other hand, \textit{OutMin} demonstrates the lowest \ac{RSNR} performance because this method considers only the outage minimization without accounting for the average \ac{RSNR}.
Meanwhile, \textit{OutMin w/ Average SE} can improve the average \ac{RSNR} close to that of \textit{LoS Concentration}, while significantly reducing the outage probability, as shown in \figref{fig:TargetSE_OutProb}.

\par
When comparing Fig.~\ref{fig:TxSNR_AverageRSNR_TargetSE=4} and Fig.~\ref{fig:TxSNR_AverageSE_TargetSE=4}, \textit{LoS Concentration} outperforms \textit{OutMin w/ Average SE} in terms of the average \ac{RSNR} but underperforms in terms of the average SE.
As described in Section~\ref{subsec:average_SE}, \textit{LoS Concentration} is designed to maximize the upper bound of the average SE, derived from Jensen's inequality in \eqref{eq:average_SE_Jensen}, which is equivalent to maximizing the average \ac{RSNR}.
Due to the maximization of the upper bound instead of directly maximizing the average SE, \textit{LoS Concentration} is inferior to \textit{OutMin w/ Average SE} in the average \ac{SE} performance.
\textit{OutMin w/ Average SE} can improve the average \ac{SE} performance, even when the LoS path is blocked, by utilizing multiple paths. 
Consequently, designing multiple beams across multiple paths enables improving the \ac{SE} while reducing outage probability owing to spatial diversity.

From these simulation results and theoretical analysis, the effectiveness of the proposed analog beamforming designs is validated in terms of achieving spatial diversity and beam gain under various target SE and SNR values.

\section{Conclusion}\label{sec:conclusion}
This paper proposed the analog beamforming designs with multi-panel arrays to enable stable \ac{mmWave} communications under stochastic blockages.
We provided the theoretical analysis of \ac{SE} and formulated the outage probability minimization problem for panel allocation.

\par
The theoretical analysis derived closed-form expressions for the \ac{CDF} and outage probability of \ac{SE}, as well as the \ac{RSNR}, for multi-beam design with multi-panel arrays.
To optimize panel allocation, two analog beamforming algorithms were proposed: one focused on minimizing outage probability and the other jointly optimizing outage probability and average \ac{SE}.
The panel allocation problem was formulated as a combinatorial optimization problem that determines the assignment of panels to both \ac{LoS} and \ac{NLoS} paths.
The proposed algorithm relies only on long-term statistics, without instantaneous \ac{CSI}.
Thus, the optimization does not need to be solved at every channel coherence time, simplifying implementation.

\par
The effectiveness of the theoretical analysis and the proposed beamforming designs was validated through computational simulations using realistic \ac{mmWave} parameters.
The results demonstrated that conventional LoS-concentrated beamforming, while optimal for maximizing average \ac{RSNR}, frequently degrades communication quality.
In contrast, the proposed multi-beam approach, which utilizes both LoS and NLoS paths, effectively suppresses outage probability and ensures more stable communication compared to conventional methods.
From these results, the proposed multi-wide-beam design achieves stable communication over stochastic blockages, outperforming the conventional single sharp-beam design.

% Future work
Finally, several directions remain open for future research.
First, beamforming design with imperfect \ac{CSI} should be investigated, since channel estimation errors may impact the robustness of the proposed panel allocation strategy.
Second, with the advent of extremely large-scale antenna arrays in future systems, the development of low-complexity algorithms will be essential for scalability.
Moreover, panel allocation strategies that explicitly account for near-field effects in the large-array systems represent a key avenue for further exploration.

%%%%%%%%%% Appendix %%%%%%%%%%
{\appendices
\section{Derivation of the Equivalent Channel Response}\label{app:heq}

We derive the \ac{PDF} of the equivalent channel response $h_\mathrm{eq}$ in \eqref{eq:PDF_heq}. 
According to the approximation in \eqref{eq:heq_approx}, the equivalent channel response is represented as the sum of $g_{q_l} \tilde{\omega}_l$, where $g_{q_l}$ and  $\tilde{\omega}_l$ denote the circularly symmetric complex Gaussian and Bernoulli random variables, respectively.
Since $g_{q_l} \tilde{\omega}_l$ is also expressed as a Bernoulli-Gaussian random variable, $h_\mathrm{eq}$ can be represented as a Bernoulli-Gaussian mixture random variable.
The equivalent channel response $h_\mathrm{eq}$ in \eqref{eq:heq_approx} can be reformulated as
\begin{align}\label{eq:heq_appendix}
    h_\mathrm{eq} = \dfrac{N_\mathrm{a}}{\sqrt{N_\mathrm{t}}} \sum_{l\in\mathcal{L}} g_{q_l} \tilde{\omega}_l =  \dfrac{N_\mathrm{a}}{\sqrt{N_\mathrm{t}}} {\sum_{l\in\mathcal{L}} e_l},
\end{align}
where $e_l$ is a dominant component of the equivalent channel response at the $l$-th path and it follows the Bernoulli-Gaussian distribution.
Here, the \acp{PDF} of $g_{q_l}$, $\tilde{\omega}_l$ and $e_l$ are defined as
\begin{align}
    \label{eq:gaussian_PDF}
    f_{G_{q_l}}(g_{q_l}) &= \mathcal{CN}(g_{q_l}; 0, \rho_l^2) = \dfrac{1}{\pi \rho_l^2} \mathrm{exp}\left(-\dfrac{|g_{q_l}|^2}{\rho_l^2}\right),\\
    \label{eq:bernoulli_PDF}
    f_{\tilde{\Omega_l}}(\tilde{\omega}_l) &
    = \mathcal{B}(\tilde{\omega}_l; p_\mathrm{blk}) = p_\mathrm{blk}^{1-\tilde{\omega}_l} (1 - p_\mathrm{blk})^{\tilde{\omega}_l},
    \\
    \label{eq:gaussian_bernoulli_PDF}
    f_{E_l}(e_l) &= p_\mathrm{blk}\delta(e_l) + (1-p_\mathrm{blk}) f_{G_{q_l}}(e_{l}).
\end{align}

\par
According to the equation \eqref{eq:heq_appendix}, $h_\mathrm{eq}$ is represented as the sum of Bernoulli-Gaussian random variables $e_l$ and follows the Bernoulli-Gaussian mixture distribution.
Since $e_l$ is independent for each $l$, the \ac{PDF} of $h_\mathrm{eq}$ can be calculated as the convolution product of the \acp{PDF} of each equivalent response $e_l$.
Thus, the \ac{PDF} of $h_\mathrm{eq}$ can be written as
\begin{align}\label{eq:gaussian_bernoulli_mixture_PDF}
    f_{H_\mathrm{eq}} (h_\mathrm{eq}) = f_{E_1}(e_1) \circledast f_{E_2}(e_2) \circledast \dots \circledast f_{E_L}(e_L).
\end{align}

\par
Furthermore, by simplifying the equation \eqref{eq:gaussian_bernoulli_mixture_PDF}, we obtain the \ac{PDF} of the equivalent channel response as in \eqref{eq:PDF_heq}.

\section{Derivation of the RSNR Distribution}\label{app:RSNR}
We derive the \ac{PDF} and \ac{CDF} of the \ac{RSNR} written in \eqref{eq:PDF_totalSNR},~\eqref{eq:CDF_totalSNR}.
As shown in \eqref{eq:totalSNR_approx}, the \ac{RSNR} $\gamma$ is expressed as the square of the equivalent channel response $|h_\mathrm{eq}|^2$, which corresponds to the square of a Bernoulli-Gaussian mixture random variable.
In the following, the objective is to derive the \ac{PDF} of the \ac{RSNR}, \textit{i.e.,} the distribution of the square of a Bernoulli-Gaussian mixture random variable.

The mixture distribution shown in \eqref{eq:PDF_heq} is expressed as the sum of a $\delta$-function term derived from the Bernoulli distribution and multiple complex Gaussian distributions, represented as a weighted sum with respect to the power of $p_\mathrm{blk}$.
As each distribution does not have cross terms, the squared distribution can be calculated separately for each distribution.
First, the squared distribution of the complex Gaussian distribution is calculated. Let $x \sim \mathcal{CN}(0, \rho^2)$ be a Gaussian random variable, and its \ac{PDF} is expressed as
\begin{align}
    f_X(x) = \mathcal{CN}(x; 0, \rho^2) = \dfrac{1}{\pi\rho^2}\mathrm{exp}\left(-\dfrac{|x|^2}{\rho^2}\right).
\end{align}

\par
Let $x$ be defined as $x = x_\mathrm{r} + j x_\mathrm{i}$, where $x_\mathrm{r} = r\cos{\theta}$ and $x_\mathrm{i} = r\sin{\theta}$.
Using this variable transformation, the \ac{PDF} of the random variable $y = |x|^2 \geq 0$ is expressed as
\begin{align}
    f_Y(y) &= \int_0^{2\pi} \int_{0}^\infty f_X(r,\theta) \delta (y - r^2) rdrd\theta, \nonumber \\
    &= \int_0^{2\pi} \int_{0}^\infty \dfrac{1}{\pi\rho^2}\mathrm{exp}\left(-\dfrac{r^2}{\rho^2}\right) \delta(y - r^2) rdrd\theta, \nonumber \\
    &= \int_0^{2\pi} \int_{0}^\infty \dfrac{1}{\pi\rho^2}\mathrm{exp}\left(-\dfrac{r^2}{\rho^2}\right) \dfrac{1}{2\sqrt{y}} \delta(r - \sqrt{y}) rdrd\theta, \nonumber \\
    &= \int_0^{2\pi} \dfrac{1}{\pi\rho^2}\mathrm{exp}\left(-\dfrac{y}{\rho^2}\right) \dfrac{1}{2\sqrt{y}} \sqrt{y} d\theta, \nonumber \\
    &= \dfrac{1}{\pi\rho^2}\mathrm{exp}\left(-\dfrac{y}{\rho^2}\right)\dfrac{\sqrt{y}}{2\sqrt{y}} \cdot 2\pi, \nonumber \\
    &= \dfrac{1}{\rho^2}\mathrm{exp}\left(-\dfrac{y}{\rho^2}\right),
\end{align}
where the property of the $\delta$-function is utilized, and the transformation $\delta(y - r^2) = \dfrac{1}{2\sqrt{y}} \delta(r - \sqrt{y})$ is applied.
Therefore, the Gaussian distribution component can be expressed as a weighted sum of exponential distributions.
Additionally, for the $\delta$-function component in \eqref{eq:PDF_heq}, squaring it still results in a $\delta$-function.
Consequently, the \ac{PDF} of the \ac{RSNR} $\gamma$, which follows the squared Bernoulli-Gaussian mixture distribution, can be expressed as shown in \eqref{eq:PDF_totalSNR}.

\par
Subsequently, we rewrite the \ac{PDF} shown in \eqref{eq:PDF_totalSNR} as a \ac{CDF}.
This can also be determined by independently calculating the \ac{CDF} for each distribution and then computing their weighted sum, which can be expressed as \eqref{eq:CDF_totalSNR}.

}

\bibliographystyle{IEEEtran}
{\bibliography{ref.bib}}
% \end{thebibliography}

%%%%% BIOGRAPHY %%%%%
%%% \newpage必要？ %%%
% \newpage
% \input{biography/bio}

\vfill

\end{document}